\documentclass[a4paper, reprint, aps]{revtex4-2}

\usepackage[utf8]{inputenc} 
\usepackage[T1]{fontenc}    
\usepackage{hyperref}       
\usepackage{url}            
\usepackage{booktabs}       
\usepackage{amsfonts}       
\usepackage{nicefrac}       
\usepackage{microtype}      
\usepackage{graphicx}
\usepackage{amsmath}

\usepackage{xcolor}

\usepackage[normalem]{ulem}

\begin{document}

\title{Detecting Hidden Layers from Spreading Dynamics on Complex Networks}

\author{Łukasz G. Gajewski}
\email{lukaszgajewski@tuta.io}
\author{Jan Chołoniewski}
\affiliation{Center of Excellence for Complex Systems Research, Faculty of Physics, Warsaw University of Technology, Koszykowa 75, 00-662, Warsaw, Poland}
\author{Mateusz Wilinski}
\affiliation{Theoretical Division, Los Alamos National Laboratory, Los Alamos, New Mexico 87545, USA}

\begin{abstract}
    When dealing with spreading processes on networks it can be of the utmost importance to test the reliability of data and identify potential unobserved spreading paths.
    In this paper we address these problems and propose methods for hidden layer identification and reconstruction.
    We also explore the interplay between difficulty of the task and the structure of the multilayer network describing the whole system where the spreading process occurs.
    Our methods stem from an exact expression for the likelihood of a cascade in the Susceptible-Infected model on an arbitrary graph.
    We then show that by imploring statistical properties of unimodal distributions and simple heuristics describing joint likelihood of a series of cascades one can obtain an estimate of both existence of a hidden layer and its content with success rates far exceeding those of a null model.
    We conduct our analyses on both synthetic and real-world networks providing evidence for the viability of the approach presented.
\end{abstract}

\maketitle

\section{Introduction}

\begin{figure*}[tb]
    \centering
    \includegraphics[width=0.95\textwidth]{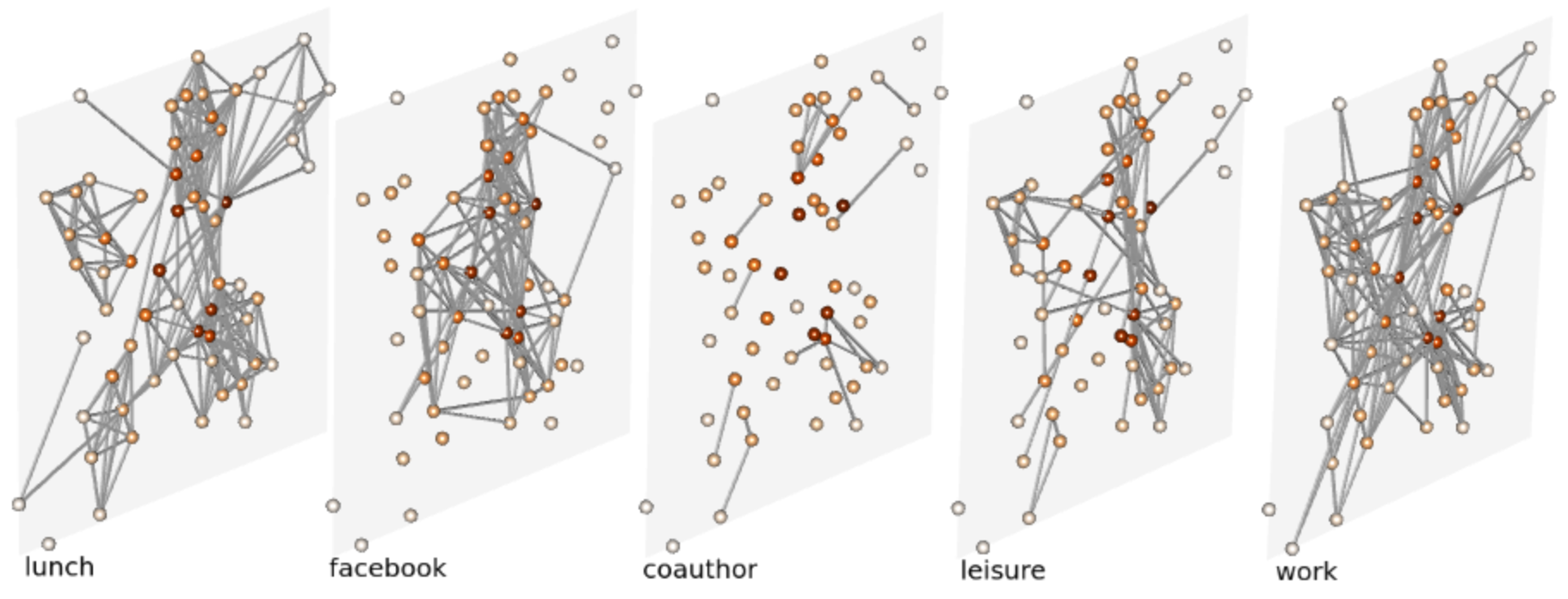}
    \caption{Visualisation of the multilayer network representing the Aarhus data \cite{magnani2013combinatorial}. We utilise this network as a real world example of possible application of our methods. It is quite natural to imagine we know only one of the layers presented here and would like to infer the existence and possibly structure of others. Note that co-author and leisure layers have lower connections' density in comparison with others. Thus we do not use them as visible layers in our analyses as that makes the task of detecting hidden connections potentially much easier. }
    \label{fig:aarhus_struct}
\end{figure*}

Real-world complex systems can often be described by interconnected structures known as multilayer networks \cite{de2013mathematical,kivela2014multilayer,boccaletti2014structure, de2015ranking}.
Transportation, social or economic networks, to name just a few general examples, can have various types of connections, see Fig.~\ref{fig:aarhus_struct} for an example depiction of such a system.
Each such type of a connection in a network can be represented as a specific sub-system or sub-network.
Railway, flights and bus connections can all be described with a network but to have a full description of the transportation system, they need to be joined and described with a multilayer network.
In reality, obtaining full information which would allow to create a complete multilayer network is rarely possible.
Moreover, in some cases, even the knowledge about all existing layers is limited.
As a result, researchers often have to deal with uncertainty which arise from dealing with partial information about connectivity in analysed system.
This specifically concerns one of the fundamental problems in network science, the spreading processes on networks \cite{barrat2008dynamical,pastor2015epidemic,de2016physics,de2018fundamentals, paluch2020source, gomez2013diffusion, sole2013spectral} but is also of significance for opinion dynamics \cite{chmiel2015phase, chmiel2017tricriticality, chmiel2020veritable, gajewski2021bifurcations}.

In the following article we focus on the problem of detecting hidden layers based on observations of a dynamical processes on graphs.
By dynamical process we mean a realisation of a spreading process described by a model of the susceptible-infected-recovered (SIR) type.
Note that such models can describe not only infections, but also spreading of information, opinions or failures.
Furthermore, we assume that the observation of such a process is limited to the states of the nodes, without the knowledge of the actual spreading path.
In the rest of the text we will refer to a single spreading realisation as a cascade.
We also propose and explore methods for finding missing connections of different types.
Finally, we analyse potential limitations and difficulties as well as beneficial settings, i.e. when solving the problem is easier, for these methods.

The problem of detecting hidden layers has appeared recently in the literature in a non-markovian setting \cite{lacasa2018multiplex} and on quantum graphs \cite{gajewski2020discovering}.
It is also closely related to the problem of network reconstruction which was extensively analysed in the past \cite{gomez2012inferring,abrahao2013trace,pouget2015inferring,braunstein2019network, netrapalli2012learning, braunstein2019network} and also in a partial observation setting \cite{lokhov2016reconstructing,woo2020iterative,wilinski2020scalable}.
Our setting is a bit simpler in some regards but at the same time still fairly realistic and thus should still be viable for real-world problems.
We feel that our simplifications are justified since solving the general problem was proved to be limited \cite{abrahao2013trace} and previous papers often approached only limited cases anyway, such as very short cascades \cite{gripon2013reconstructing}.
This is not to say that successful approximations are not possible \cite{gomez2012inferring}, however, the goal of this paper is to investigate the challenges associated with detecting hidden layers in interconnected networks in the context of spreading processes.

Reader should not confuse the problem of finding hidden layers based on observed spreading with extensively analysed branch of network science called \textit{link prediction}, where the hidden connections are estimated using only the network structure.
A seminal paper in this direction is \cite{guimera2009missing}.
An extension, including multilayer networks, can be found in \cite{de2017community}.

The paper is structured as follows: in the next section we describe all the methods used in the analysis, starting with tools which allow for detecting hidden layers and then proposing methods for identifying unobserved connections.
In the third section we analyse both synthetic and real world networks and show how our methods work under different structural circumstances.
Finally, we discuss all the results and present our conclusions in the last section.

\section{Methods}

In our analysis we focus on one of the simplest spreading model -- the Susceptible Infected (SI) model \cite{hurley2006basic}.
We use it in a network version where the dynamics can be described as follows: for each node $i$, which is in the infected state $I$ at time $t$, each of its neighbors $j$ (in susceptible state $S$) will become infected at time $t+1$ with probability $\beta$.
This model was used because of its simplicity on one hand and the mechanism of multiple infection opportunities (in comparison with, e.g. an Independent Cascade model) on the other.
The latter makes the combinatorial analysis much more difficult, as we will see in further sections.
The former is reflected by an easy to derive likelihood of any observed spreading, including a multilayer scenario, assuming that the full knowledge about connectivity is available.
We are well aware that for specific applications other, often more complex, models may be better suited for the problem.
Our approach, however, can effectively be used for other spreading description.
As an example, we provide the full derivation and results also for the Independent Cascade model (see Appendix \ref{sec:app_ic}).
Note that SI and IC are limiting versions of the SIR model, which means that the presented results give limits for the whole family of SIR models.

Let us denote a multilayer graph with $G$ and let the probability of infection (spreading) on each layer $j$ be equal to $\beta_j$.
We will refer to a single spreading dynamics as a cascade and denote it with $\Sigma^c$.
A single cascade can be described by a set of infection times $\tau_i^c$ for each node $i$.
We will also assume that a cascade ends at time $t_{max}$ and if a given node was not infected at all, its infection time will be equal to $t_{max}$.
In other words, if node's $i$ activation times is equal to $t_{max}$, it was either activated at $t_{max}$ or later -- this will be more clear once the likelihood is derived.
The set of all available cascades will be denoted with $\Sigma$.

\subsection{Cascade likelihood for the Susceptible-Infected model}

As mentioned before, the likelihood of a given set of cascades, for a specific and fully known multilayer network can be derived, similarly as it was done in \cite{lokhov2016reconstructing}.
In short, the probability of a given data-set can be written as a product over cascades, which are independent, and nodes, because the problem can be considered locally:
\begin{equation}
    P(\Sigma | G, \{ \beta_j \}) = \prod_{i \in V} \prod_{c \in C} P_i(\tau_i^c | \Sigma^c, G, \{ \beta_j \}),
    \label{eq:likelihood}
\end{equation}
where $V$ is the set of nodes in graph $G$.
Note that we can use our knowledge about the activation times of each node and compute the probability of node $i$ not being activated by its neighbor $k$ (from layer $j$) before $\tau_i^c - 1$ under cascade $c$, which is equal to $\prod_{t=0}^{\tau_i^c - 2} (1 - \beta_j \mathbf{1}_{\tau_k^c \leq t})$, where $\mathbf{1}_{x}$ is the indicator function (it is equal to $1$ when $x$ is true).
Similarly, the probability of the same node not being activated by its neighbor exactly at time $\tau_i^c - 1$ is equal to $(1 - \beta_j \mathbf{1}_{\tau_k^c \leq \tau_i^c - 1})$.
Have in mind that $\tau_i^c = t_{max}$ is equivalent to node $i$ not being activated at all.
Then each element of the product in Eq. (\ref{eq:likelihood}), being the probability of node activation under a specific cascade, can be computed as follows:
\begin{equation}
    \begin{split}
        &P_i(\tau_i^c | \Sigma^c, G, \{ \beta_j \}) = \Bigg(\prod_{t=0}^{\tau_i^c - 2} \prod_j \prod_{k \in \partial_j i} (1 - \beta_j \mathbf{1}_{\tau_k^c \leq t}) \Bigg) \\
        &\times \Bigg( 1 - \prod_j \prod_{k \in \partial_j i} (1 - \beta_j \mathbf{1}_{\tau_k^c \leq \tau_i^c - 1}) \mathbf{1}_{\tau_i^c < t_{max}} \Bigg),
    \end{split}
    \label{eq:local_likelihood}
\end{equation}
where $\partial_j i$ is the set of neighbors of node $i$ in layer $j$.
If we know one or more layers of the network, we can use the above equation to compute the probability of observed dynamics, assuming that there are no other propagation channels.
As seen from Eq. (\ref{eq:likelihood}), adding more cascades reduces the likelihood.
Nevertheless increased statistics of cascades makes it easier to find hidden edges, as it is shown in Fig.~\ref{fig:scal}.

\subsection{Detecting the existence of a hidden layer}
An unobserved layer of propagation may lead to a situation where nodes become active despite not interacting, on the observed network, with any other active node.
Such a situation leads to the probability described by Eq. (\ref{eq:likelihood}) being equal to zero.
In such a case, one can be sure that there exist spreading paths that are not present in the observed layers.
In reality, different layers share certain similarities and may be strongly correlated which would decrease the probability of observing a forbidden activation -- in other words, it is less likely that an activation, which occur through a hidden connection, will have a zero probability.
Real-life social networks are a good example of this because of a significant overlap between connections on different social media platforms.
Moreover, social connectivity is characterised by high clustering \cite{white1976social} which increases the probability of observed neighborhood being active, even though the activation came through an unobserved edge.
To address such a situation and still be able to evaluate whether a given set of cascades indicates existence of an unobserved layer or not, we assume a single layer model and utilise the Vysochanskij–Petunin inequality (VP) \cite{pukelsheim1994three} to evaluate whether a given cascade could be generated by such a model.

We expect that cascades, which involved spreading through unobserved edges, have significantly lower likelihood than those in the assumed null model.
Assuming that the likelihood distribution is unimodal, we can quantify how distant the value of an observed cascade likelihood is from the value expected from the single-layer model using the VP inequality.
It states that if $X$ is a random variable with a unimodal distribution (mean $\mu$, finite and positive variance $\sigma$) and $\lambda > \sqrt{8/3}$, then:
\[
\mathrm{Pr}(|X-\mu|\geq \lambda\sigma)\leq\frac{4}{9\lambda^2},
\]
which after normalisation $\tilde{X}=\frac{|X-\mu|}{\sigma}$ gives:
\[
\mathrm{Pr}\left(\tilde{X}\geq \lambda\right)\leq\frac{4}{9\lambda^2}.
\]
Inserting $\lambda=\tilde{x}$:
\[
\mathrm{Pr}\left(\tilde{X}\geq \tilde{x}\right)\leq\frac{4}{9\tilde{x}^2},
\]
which means that an upper bound of probability of obtaining a result $\tilde{x}$ or greater from a normalized unimodal distribution describing $\tilde{X}$ is:
\begin{equation} 
p(\tilde{x}) = \min\left(\frac{4}{9\tilde{x}^2},1\right)
\label{eq:p_max}   
\end{equation}
for $\tilde{x}>\sqrt{8/3}$.
In our case, $x$ is the likelihood of a cascade given by Eq. (\ref{eq:likelihood}), $X$ is a random variable describing the likelihood values for an assumed single layer model.
After performing simulations, we calculate $\mu$ and $\sigma$, and normalise the $x$ value obtaining $\tilde{x}$. 

In our experience the distributions, which describe $X$ in the single-layer case are unimodal and $\tilde{x}>\sqrt{8/3}$ with a sufficient observation time ($t_{max}$), thus the assumptions of the VP hold, however, as we are the ones producing these distributions then it is trivial to inspect whether that is the case before proceeding with the rest of our method.
In the event, it is not true an alternative theorem, analogous to the VP inequality (e.g. the Chebyshev's inequality \cite{saw1984chebyshev}), can be introduced or even more general approaches such as bootstrapping \cite{efron1994introduction} can be used to determine the confidence level of the observation.

In the paper, we use $p(\tilde{x})$ to measure how surprising given cascades are, assuming they were generated by an SI process with known $\beta$ simulated on a visible layer of the network. 
We validate our approach in the next section, using simulations and synthetic networks.

\subsection{Detecting the hidden edges}

In the previous section, we shown how one can discard the possibility that observed cascades were generated by the given network.
Once the existence of the hidden layer is established, the same data can be used to estimate the topology of a hidden layer(s).
This requires assuming some topology (given visible layer and estimated hidden layer) and $\beta_{hidden}$ to simulate the process again and calculate new likelihoods.
We will try to find the topology by finding cases of node activation that could not be explained by the assumed single layer model.
Then for each such activation we identify the set of potential hidden edges.
The details of the procedure are as follows:
\begin{enumerate}
    \item Let $\mathcal{J}^c(t)=\{i\in \mathcal{N}:\tau^c_i\leq t\}$ be a set of nodes that \textit{are infected} in a simulation step $t$ of cascade $c$. $\mathcal{N}$ represents the set of nodes in graph $G$ and $\tau^c_i$ is the activation time of node $i$ in cascade $c$. $\Delta\mathcal{J}^c(t)=\left\{i\in \mathcal{N}:\tau^c_i= t\right\}$ will be a set of nodes that \textit{became infected} in a simulation step $t$ of cascade $c$.
    \item Using above notation we can introduce a set of nodes that became infected in simulation step $t$ of cascade $c$ but were not a neighbor of any node infected at $t-1$:
    \[
    \mathcal{U}^c(t)=\Delta\mathcal{J}^c(t) \setminus \bigcup_{m \in \mathcal{J}^c(t-1)}\partial m,
    \]
    where $\partial m$ is the known neighborhood of node $m$.
    \item If the likelihood of a cascade is zero then there is at least one hidden edge in a set:
    \[\mathcal{E}(i,c)=\left\{(i,k):k \in \mathcal{J}^c(\tau^c_i-1), \,\,i \in \bigcup_{t=1}^{t_{max}}\mathcal{U}^c(t)\right\}.\]
    Likelihood of such an edge being the one that was activated to infect $i$ is unknown and difficult to find.
    However we can say heuristically that said likelihood is:
    \[P^c(i,k) \sim (1-\beta_{hidden})^{|\tau^c_i-\tau^c_k|}.\]
    \item Then we must unify the candidates amongst the cascades.
    Namely if an edge (a, b) was detected in $c=1$ but not in $c=2$ we need to associate it with the likelihood of \textit{not being detected} which is similarly non-trivial.
    Also similarly we can say whatever that likelihood is it must be $\sim (1-\beta_{hidden})^{|\tau^c_i-\tau^c_k|}$
    \item Finally, for each edge $e$ we multiply its likelihoods, therefore obtaining a \textit{joint likelihood} $J$:
    \[
    J(e) = \prod_c P^c(e).
    \]
    \item Edges that maximise $J$ are most likely the hidden edges we seek.
\end{enumerate}

In order to evaluate the quality of our approach, we will use two metrics:

\textbf{Sensitivity} - the ratio between true-positives and all positives. In our case it is the fraction of hidden edges that were detected. As we simulate our systems many times, we take the mean of the sensitivity across all simulations.

\textbf{$\alpha$ - Credible Set Size} ($\alpha$-CSS) - a measure introduced in \cite{paluch2020optimizing}.
It represents the number of candidates one must investigate in order to have $\alpha$ level of certainty of finding the sought entity.
In practice one computes the rank, i.e., the position, of the entity one wishes to find, in the list of the candidates, which is in a descending order, in accordance to a given measure said entity should maximise.
This is repeated many times in order to get a distribution of that rank, and finally, one takes a quantile $q=\alpha$ of that distribution, thus acquiring the $\alpha$-CSS value.
In our case the measure is the \textit{joint likelihood} $J$ described above in point 5., and there are multiple entities -- edges -- thus we have taken the liberty of adapting said measure such that we take the \textit{highest recorded rank} amongst the hidden edges and follow the rest as usual. I.e., we compute $J$ for the appropriate edges (see the procedure description 1.-6. above), order these edges by their $J(e)$ (highest to lowest), find the position of the actual hidden edges (their rank), take the highest recorded rank, repeat $10^4$ times, take the quantile $q=0.95$ of these ranks.

\begin{figure}[!htb]
    \centering
    \includegraphics[width=0.47\textwidth]{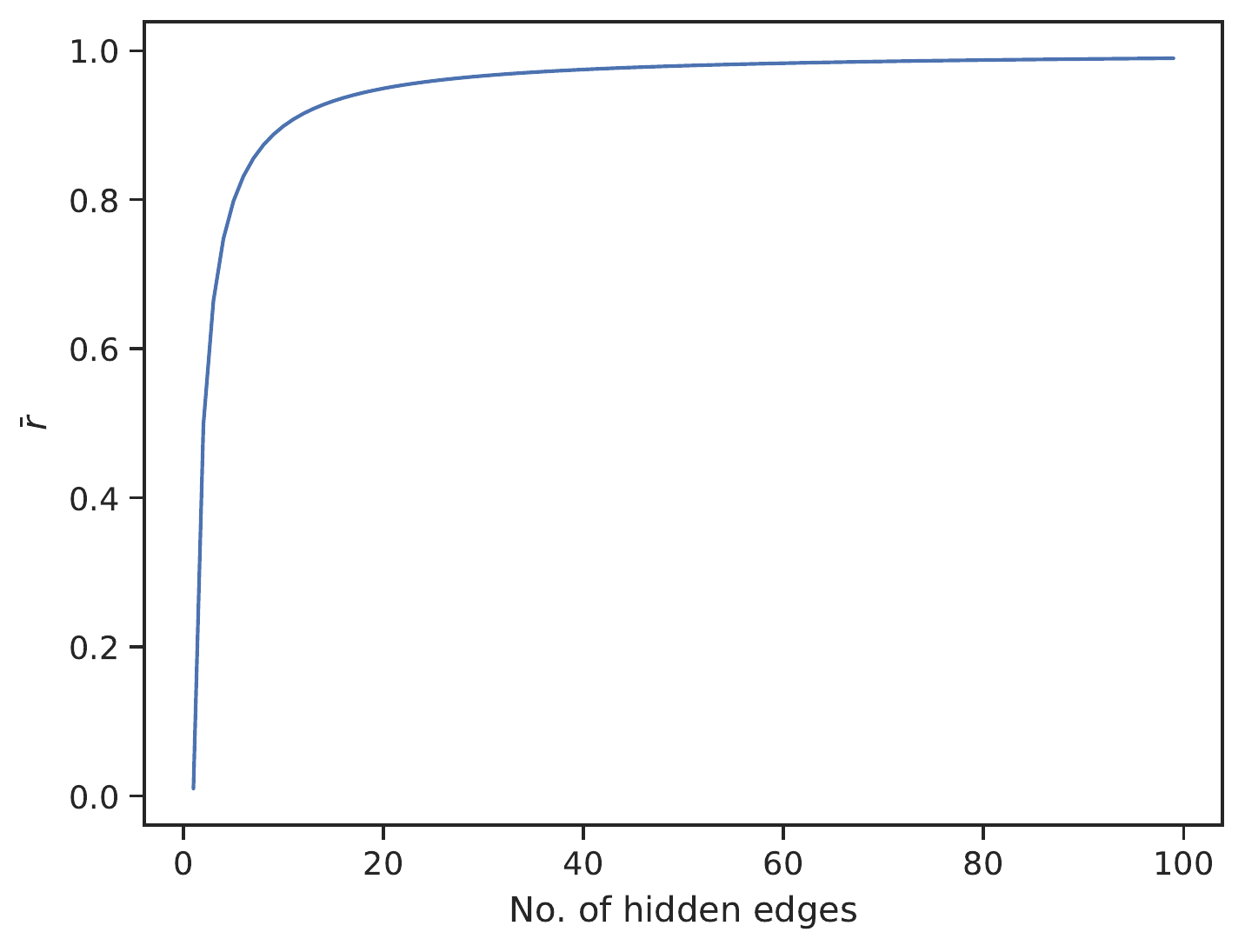}
    \caption{The ratio of edges required to check in order to have 95\% certainty, according to the null model, of testing all hidden edges, as a function of number of hidden edges.
    The plot is done for a network with $N=100$ nodes, but the shape of the curve scales with the size of the network.}
    \label{fig:null}
\end{figure}

A null model would naturally be random guessing.
There are ${N\choose2}$ edges to check in a system with $N$ nodes.
So for instance, with $N=100$ that is: ${100\choose2} = 4950$ in which case 95\% certainty of finding 1 hidden link requires checking $0.95 \times 4950 = 4703$ (rounded up) edges.
In general the number $r$ of links required to check in order to have $\alpha$ certainty, according to the null model, can be obtained from:
\begin{equation} 
\frac{{r \choose k}}{{{N \choose 2} \choose k}} = \frac{r!}{\left(r-k\right)!}\frac{\left({N \choose 2}-k\right)!}{{N \choose 2}!} = \alpha,
\label{eq:null}   
\end{equation}
where $k$ is the number of hidden edges and $N$ is the number of nodes.
Fig. \ref{fig:null} presents the normalised $\bar{r} = \frac{2r}{N(N-1)}$ as a function of $k$ in the case of $N=100$ nodes.
As we show later on our method requires substantially less edges to be checked.
Since Eq. (\ref{eq:null}) requires to be solved numerically, we also derive an asymptotic approximation, which can be find in the 
Appendix \ref{sec:app_null}, and from which we can see that $\bar{r} \sim \sqrt[k]{\alpha}$.

\section{Experiments}

We use both real and synthetic data in our experiments.
In the latter case we build networks that are realistic and not trivial in the sense that we do not want the occurrence of an activation not explained by the visible network to be likely.
To achieve that, we need a way to control the correlation between different layers of the network -- where by correlation we mean the percentage of overlapping edges.
Therefore we propose our own models for generating multilayer networks.
As for the cascades, these are in both cases generated using the SI model dynamics.
As an initial condition for each cascade we randomly pick a node and change its state into \textit{infected}.
Each cascade is generate independently.

\subsection{Synthetic Networks}

In the first setting, we generate a two-layer network using Barabási-Albert algorithm \cite{barabasi1999emergence}.
There are two parameters, which need to be selected -- $m_{hidden}$ and $m_{observed}$, which represent the number of edges added at each step of the algorithm, for hidden and observed layers respectively.
The two layers are independent but both have power-law degree distribution, which is believed to resemble real social networks \cite{clauset2009power} (although lately it is seen more as an idealised approximation \cite{broido2019scale}).
Since lack of correlation makes the problem of detecting hidden layers much easier we also propose another setting.
We take a square lattice as the first layer and then apply a rewiring procedure similar to the one introduced by Watts and Strogatz \cite{watts1998collective} to produce another layer.
We start with a square lattice as an equivalent of real relationships (affected by distance) but we also explore scale-free network as a starting layer.
The correlation between the two layers is parameterised by $p$ -- the probability of each node being rewired to any random (other) node.
In all described settings we keep the spreading probability of observed layer equal to $\beta_{observed} = 0.5$.
For the hidden layer, this probability takes the values of $\beta_{hidden} = 0.3$ and $\beta_{hidden} = 0.7$.

\begin{figure}[!htb]
    \centering
    \includegraphics[width=0.47\textwidth]{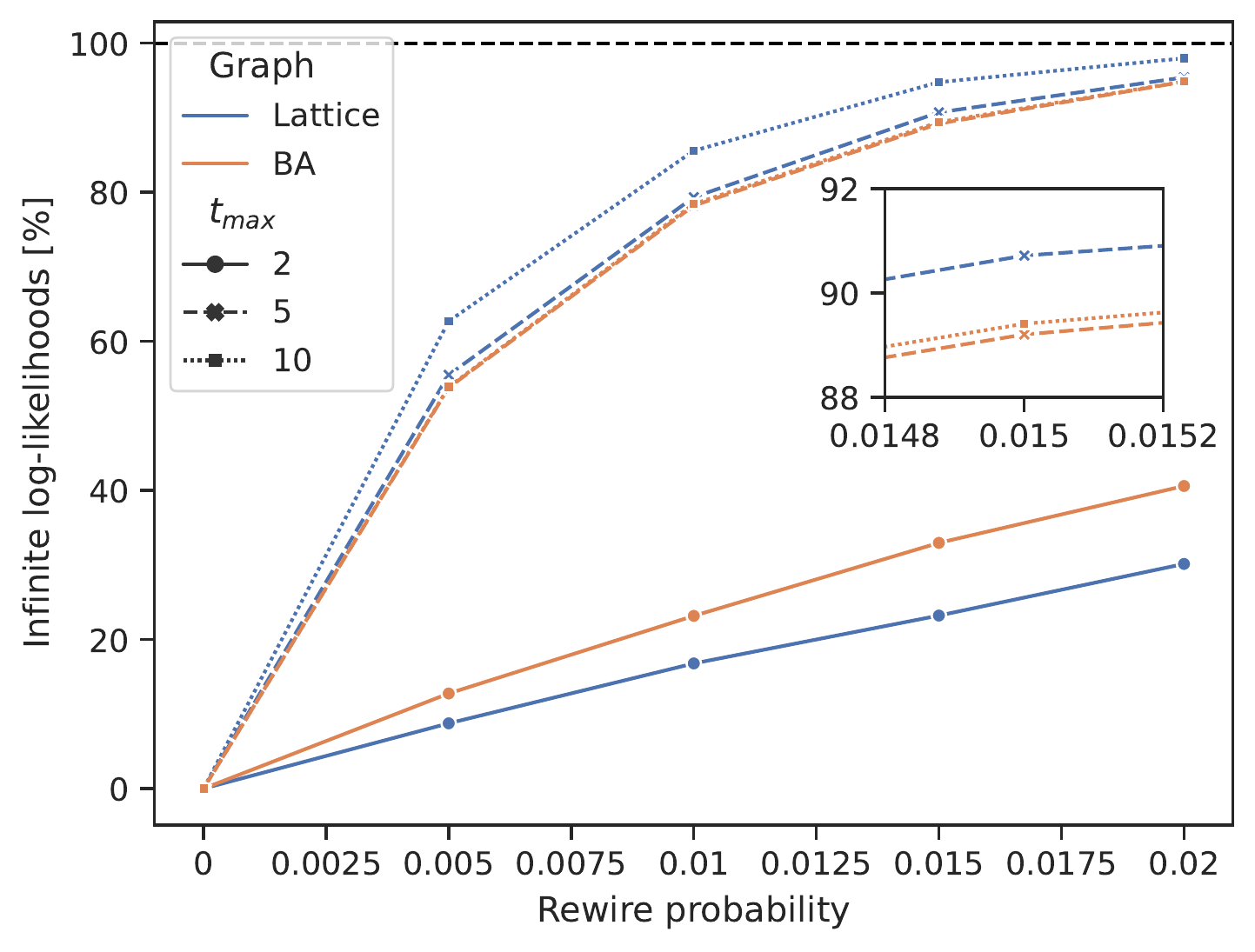}
    \caption{The percentage of log-likelihoods resulting with $-\infty$ as a function of the probability of rewiring $p$.
    We investigate two different cases of networks: a square lattice and the Barabási-Albert network for different lengths of cascades.
    The simulations were made for networks of size $N=100$ with periodic boundary conditions in the lattice case.
    The inset zooms in on the curves obtained for $t_{max} > 2$.}
    \label{fig:infinite}
\end{figure}

When it comes to detecting hidden layers using any two-layer network with independent layers results with majority of cascades giving likelihood equal to $0$.
This is a result of many non-overlapping connections, which activate nodes in a way, which would be impossible for the visible layer.
Therefore the problem becomes trivial for such setting.
The only interesting case is the model with rewiring as there the correlations between layers can be large.
Fig. \ref{fig:infinite} shows the percentage of cascades resulting with likelihood equal to zero as a function of rewiring probability.
The higher the rewiring probability, the more independent are the networks (with rewiring probability of $1$ being the fully independent case).
We test it for local networks (represented by a square lattice) and when there are many long connections (like in the case of a scale-free Barabási-Albert network).
As expected the problem is easier for short cascades and quickly becomes more difficult when the length of cascades grows.
Note that even a very small rewiring probability results in a drastic change of the discussed percentage.
It practically means that detecting an unknown transmission channel is fairly simple with our approach.
A much more challenging task, however, is to find the actual unknown connections.
We shall focus on that in further experiments but first let us discuss the case when there is no prohibited dynamics and if we can investigate the likelihood of such observed data.

\begin{figure*}[!htb]
    \centering
    \includegraphics[width=0.48\textwidth]{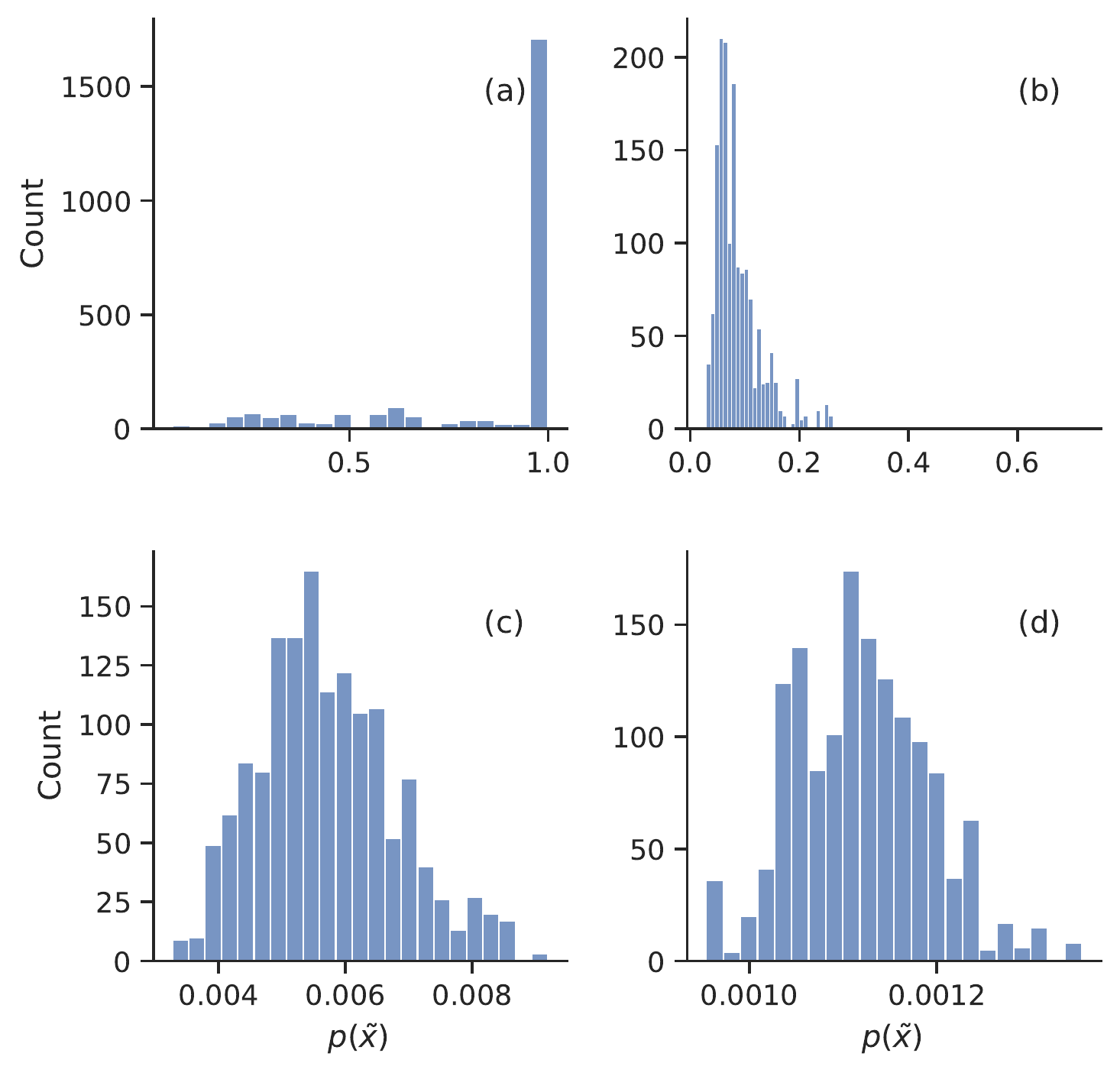}
    \hfill
    \includegraphics[width=0.465\textwidth]{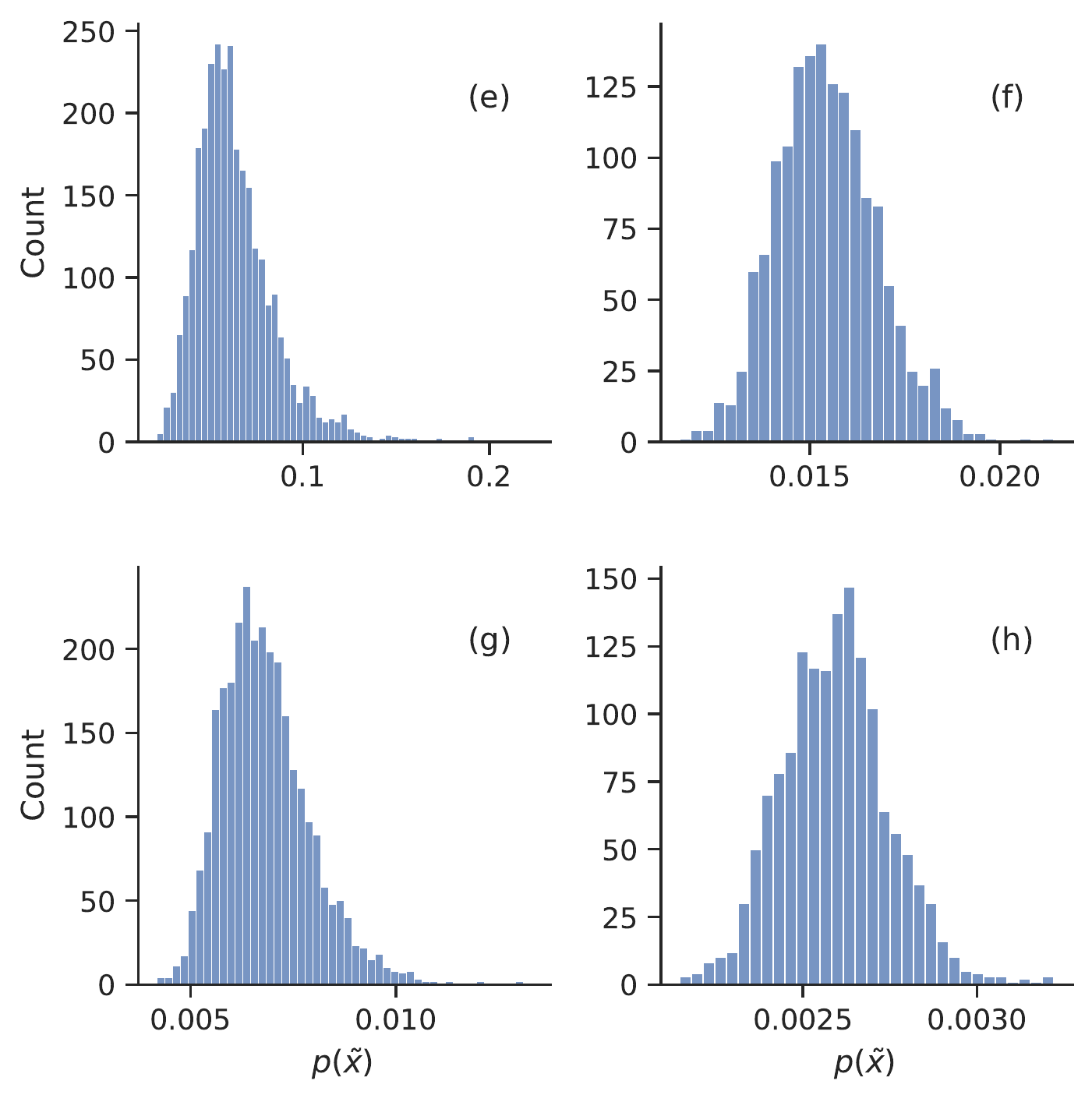}
    \caption{Histograms of $p(\tilde{x})$ -- the probability that the system has only one layer (see Eq. (\ref{eq:p_max})) -- for various combinations of $t_{max}$ and $\beta_{hidden}$.
    Plots on the left are generated for square lattice with rewiring, while the ones on the right are generated for BA network with rewiring.
    Parameters for all the networks are as follows: $\beta_{observed}=0.5$, $N=100$, $p=0.01$ and $m=3$ (in case of BA networks).
    Each histogram was made with $10^4$ realisations. 
    (a) $t_{max}=5$, $\beta_{hidden}=0.3$,
    (b) $t_{max}=5$, $\beta_{hidden}=0.7$,
    (c) $t_{max}=10$, $\beta_{hidden}=0.3$,
    (d) $t_{max}=10$, $\beta_{hidden}=0.7$,
    (e) $t_{max}=5$, $\beta_{hidden}=0.3$,
    (f) $t_{max}=5$, $\beta_{hidden}=0.7$,
    (g) $t_{max}=10$, $\beta_{hidden}=0.3$,
    (h) $t_{max}=10$, $\beta_{hidden}=0.7$,
    }
    \label{fig:p_max}
\end{figure*}

If the probability of known cascades is positive we can compare it with the empirical distribution of cascades simulated on the observed network.
This allows us to use the Vysochanskij–Petunin inequality and decide whether the observed data was generated by process run a graph with an additional (hidden from us) layer.
As seen in Fig. \ref{fig:p_max} using a typical significance level of $0.05$ allows to successfully reject the hypothesis about a single layer in significant number of cases (or even all of them).
Low $t_{max}$ and $\beta_{hidden}$ especially in the case of local networks like the square lattice decrease the effectiveness of the test but apart from the extreme case (lattice with $t_{max}=5$ and $\beta_{hidden}=0.3$) our proposed approach is an efficient tool for detecting hidden layers.

\begin{figure*}[!htb]
    \centering
    \includegraphics[width=0.48\textwidth]{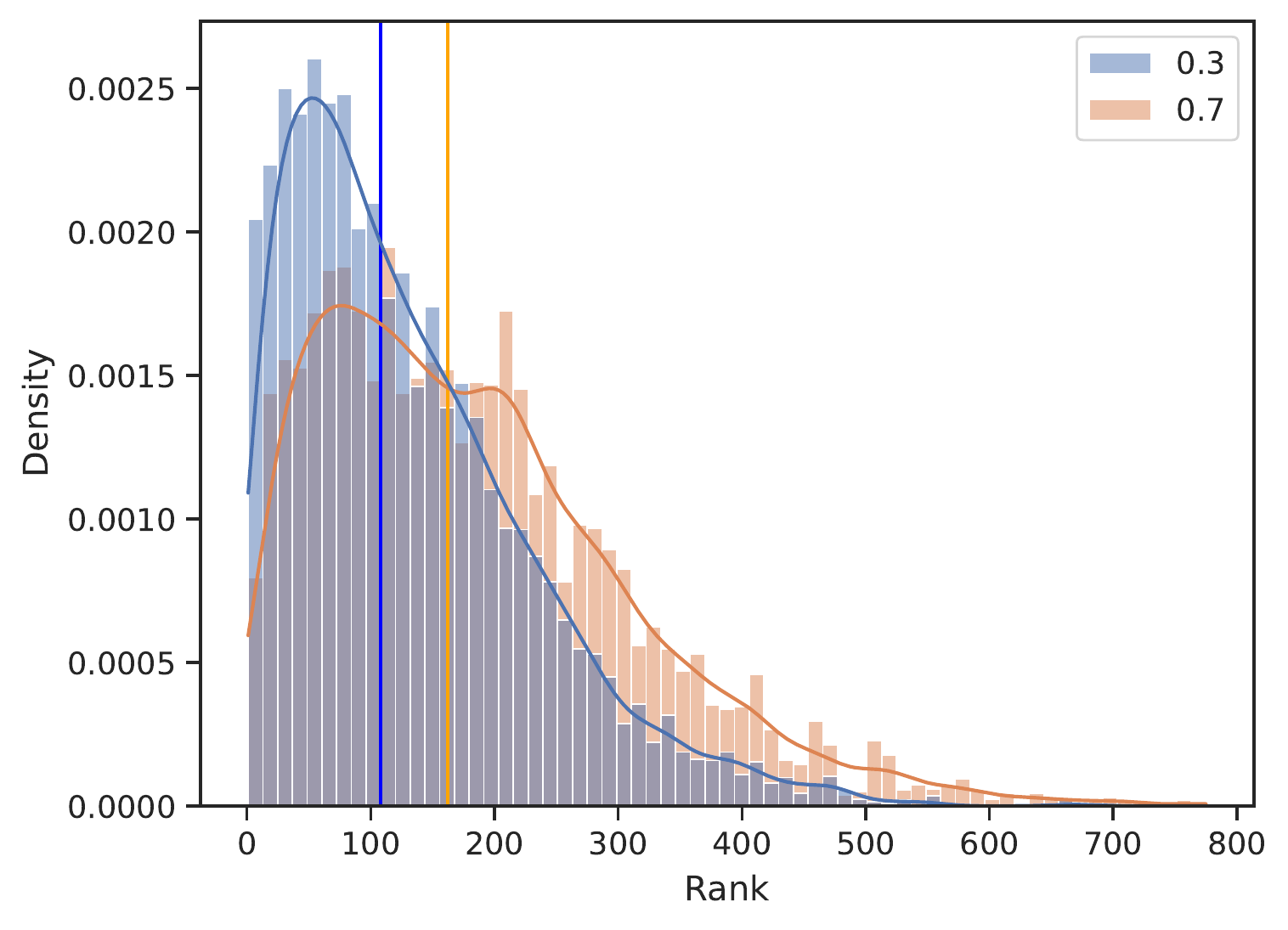}
    \hfill
    \includegraphics[width=0.47\textwidth]{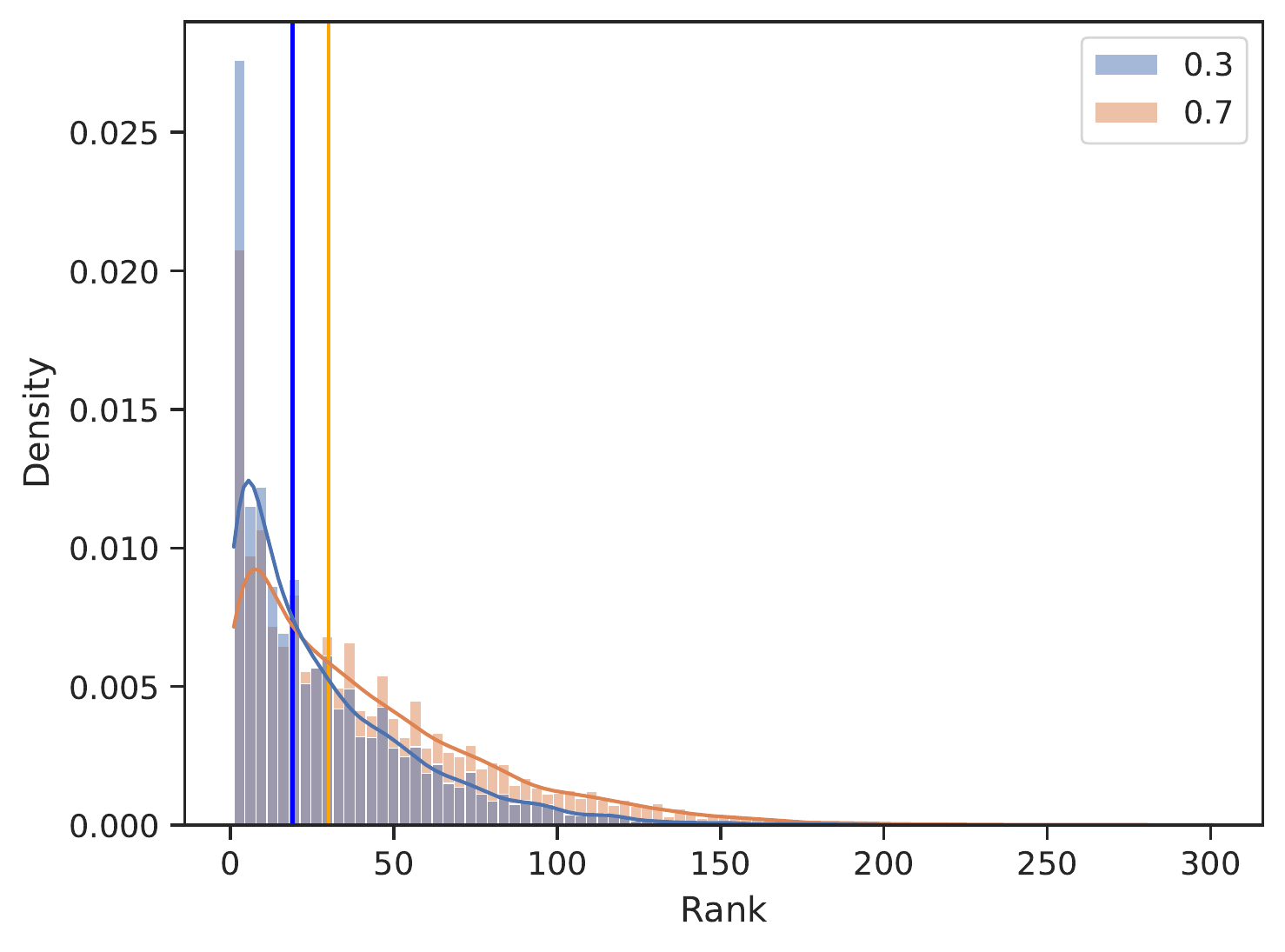}
    \caption{Distribution of ranks of hidden edges with medians as vertical lines. Left: lattice with rewiring ($N=100$, $t_{max}=10$, $p=0.01$). Right: BA network with rewiring ($N=100$, $t_{max}=10$, $m=3$, $p=0.01$).
    Results obtained with $10^4$ realisations per scenario where each scenario had $10$ independent cascades from different sources.
    These results are for those realisations where all hidden edges were detected. Solid lines show a Gaussian kernel density estimate.}
    \label{fig:ranks}
\end{figure*}

Once we know that there is a hidden layer affecting dynamics we aim at finding its edges.
Tables \ref{tab:lattice_sensitivity} and \ref{tab:ba_sensitivity} below show the results of applying our method to both lattice and Barabási-Albert networks with hidden layers produced by rewiring (with probability $p=0.01$).
When comparing the two settings one can observe a certain interplay between sensitivity and $\alpha$-CSS.
For lattice based network the sensitivity is significantly higher than in Barabási-Albert case but at the same time scale-free case is characterised by a much lower $\alpha$-CSS for both $\alpha=0.5$ and $\alpha=0.95$.
In other words, it is easier to correctly identify hidden edges when we have a locally connected network (lattice) but at the same time a scale-free network requires a smaller set to find all hidden edges (despite reaching a lower sensitivity level).
Note that in both cases the observed 0.5-CSS and 0.95-CSS are significantly lower than for the null model where, depending on the number of rewired links, they would be larger than 2475 and 4703 respectively (see Eq. (\ref{eq:null}) for $N=100$ and $k=1$ -- the higher the $k$, the more links need to be checked for the null-model).
Full distribution of ranks from which the $\alpha$-CSS was computed is shown for both networks at Fig. \ref{fig:ranks}.

\begin{table}[!htb]
\centering
\begin{tabular}{p{0.08\textwidth}p{0.08\textwidth}p{0.092\textwidth}p{0.092\textwidth}p{0.092\textwidth}}
$\beta_{hidden}$ & $\beta_{observed}$ & \textit{sensitivity} & \textit{0.5-CSS} & \textit{0.95-CSS} \\
\hline
0.3 & 0.5 & 0.81 & 108 & 322 \\
0.7 & 0.5 & 0.85 & 162 & 422
\end{tabular}
\caption{Sensitivity and $\alpha$-CSS for a square lattice with rewiring ($N=100$, $t_{max} = 10$, $p=0.01$).
Results obtained for $10^4$ realisations per scenario where each scenario had $10$ independent cascades with randomly selected sources.}
\label{tab:lattice_sensitivity}
\end{table}

\begin{table}[!htb]
\centering
\begin{tabular}{p{0.08\textwidth}p{0.08\textwidth}p{0.092\textwidth}p{0.092\textwidth}p{0.092\textwidth}}
$\beta_{hidden}$ & $\beta_{observed}$ & \textit{sensitivity} & \textit{0.5-CSS} & \textit{0.95-CSS} \\
\hline
0.3 & 0.5 & 0.53 & 19 & 87 \\
0.7 & 0.5 & 0.69 & 30 & 116
\end{tabular}
\caption{Sensitivity and $\alpha$-CSS for a Barabási-Albert network with rewiring ($N=100$, $t_{max}=10$, $m=3$, $p=0.01$).
Results obtained for $10^4$ realisations per scenario where each scenario had $10$ independent cascades with randomly selected sources.}
\label{tab:ba_sensitivity}
\end{table}

\begin{figure}[!htb]
    \centering
        \includegraphics[width=0.5\textwidth]{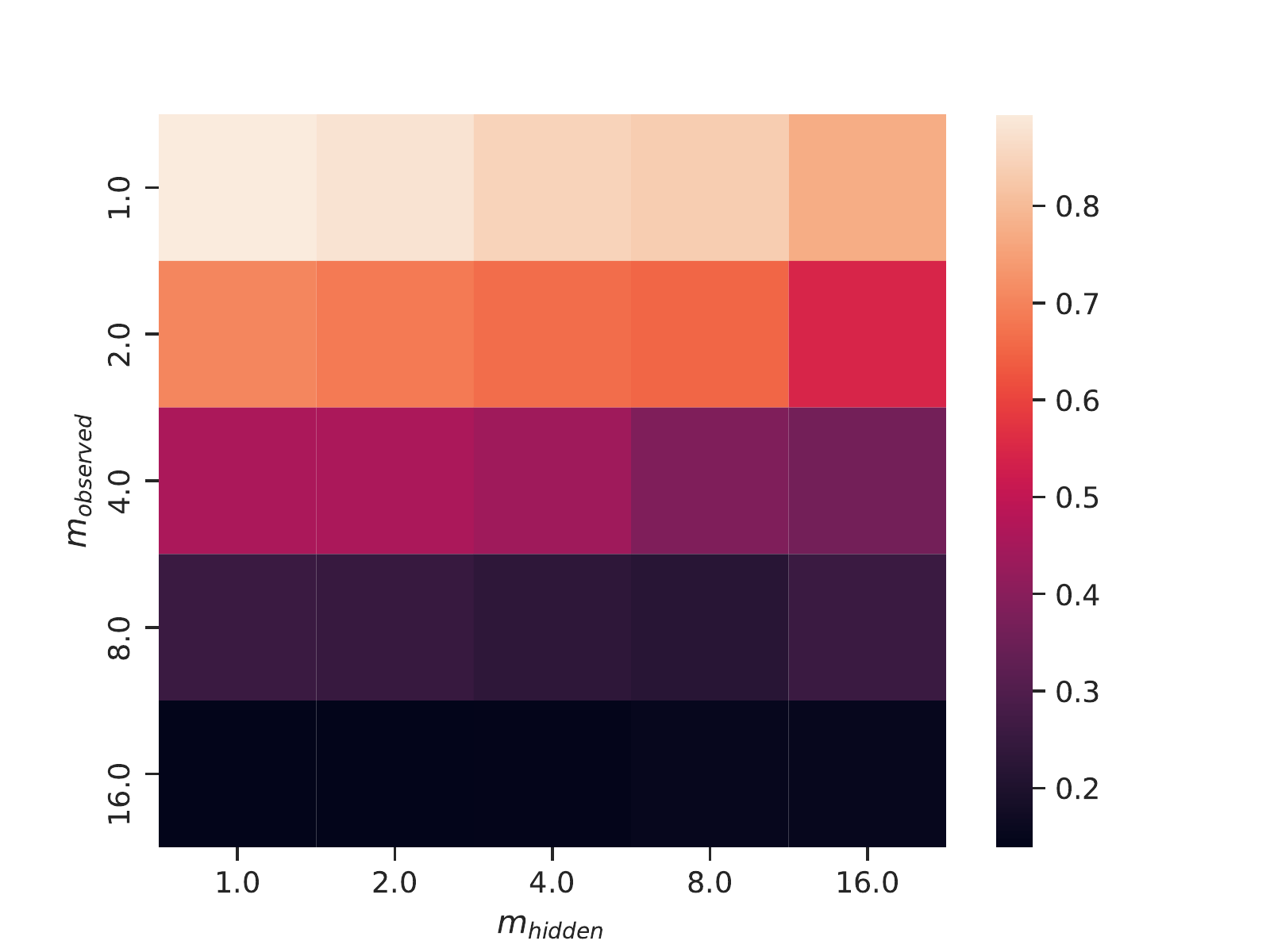}
    \vfill
        \includegraphics[width=0.5\textwidth]{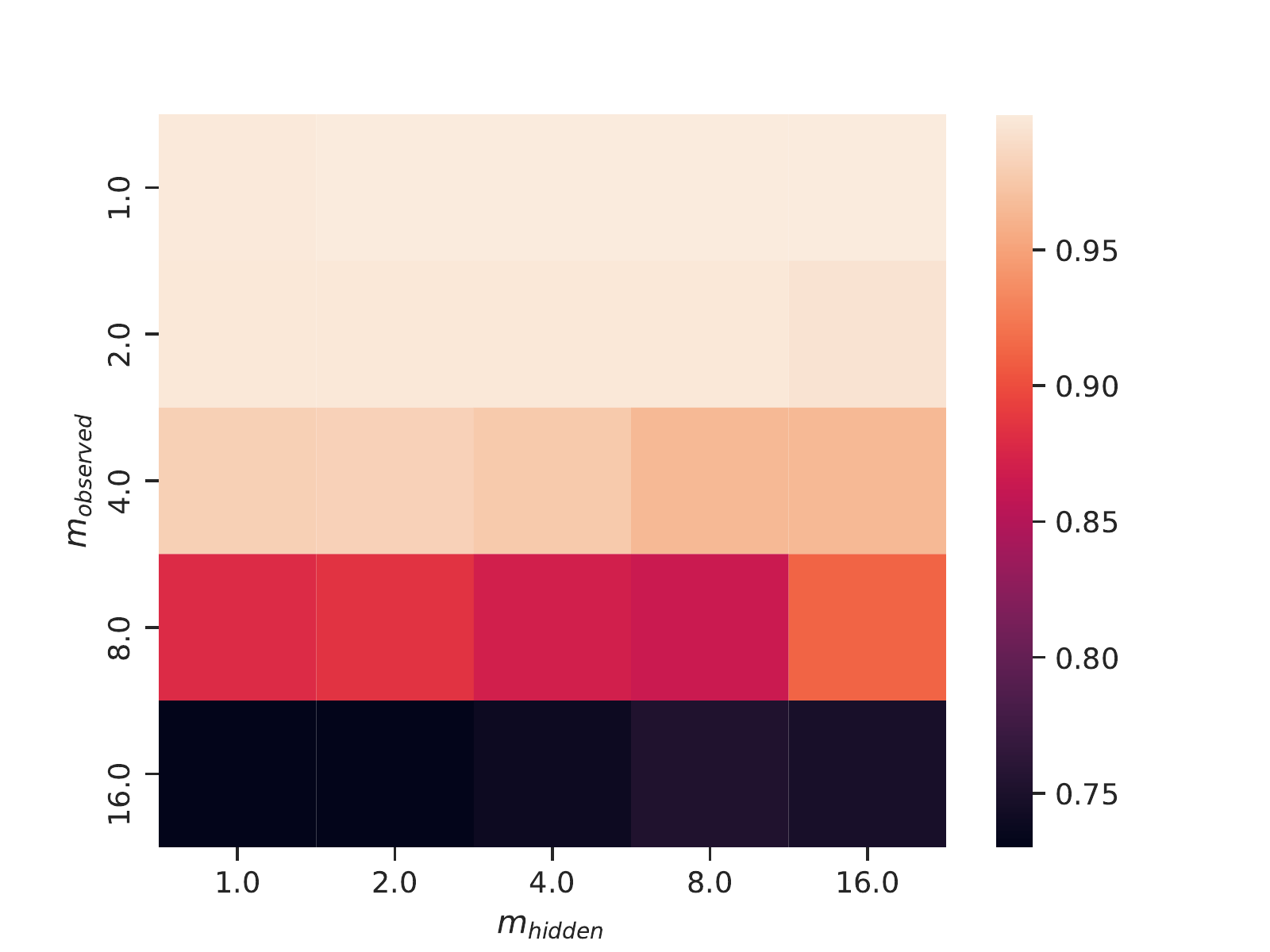}
    \caption{The sensitivity as a function of $m_{hidden}$ and $m_{observed}$ for two layer Barabási-Albert network with $\beta_{hidden}=0.7$, $\beta_{observed}=0.5$ and $t_{max}=10$ after $10$ (top) and $100$ (bottom) cascades.
    The results are averaged over 20 independent runs.}
    \label{fig:heat_sens}
\end{figure}

\begin{figure}[!htb]
    \centering
        \includegraphics[width=0.5\textwidth]{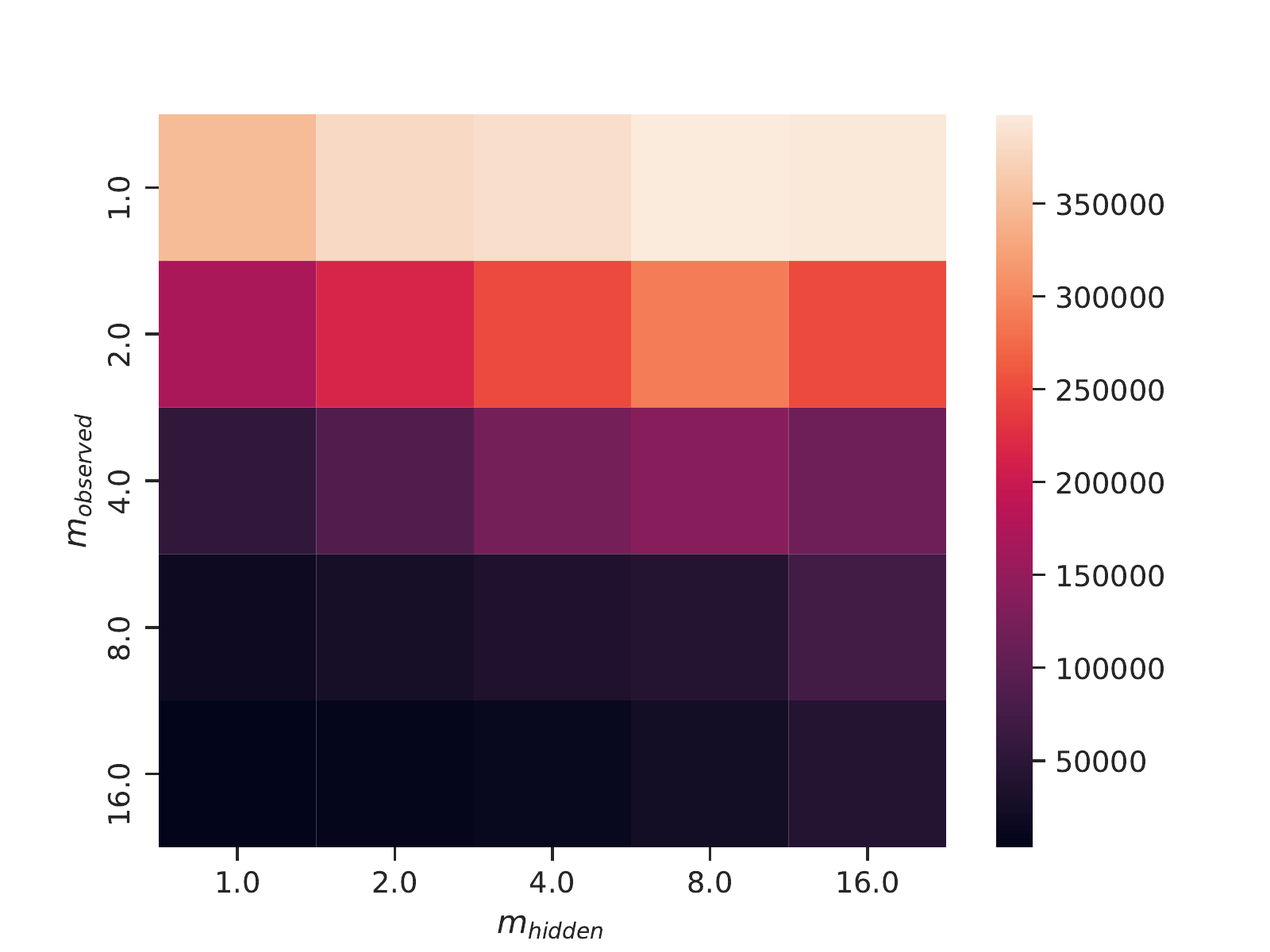}
    \vfill
        \includegraphics[width=0.5\textwidth]{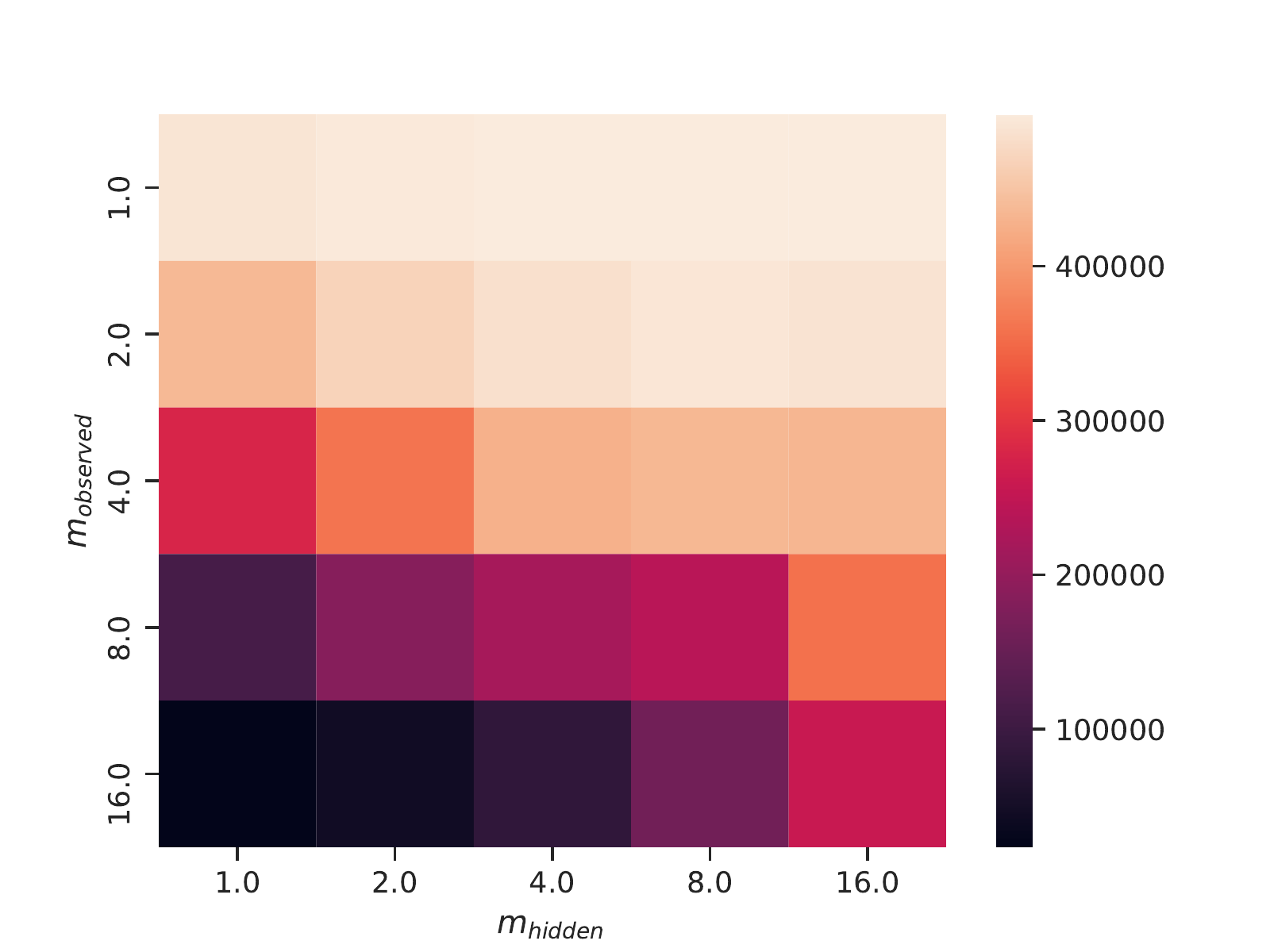}
    \caption{The 0.95-CSS as a function of $m_{hidden}$ and $m_{observed}$ for two layer Barabási-Albert network $N=1000$ with $\beta_{hidden}=0.7$, $\beta_{observed}=0.5$ and $t_{max}=10$ after $10$ (top) and $100$ (bottom) cascades.
    The results are from 20 independent runs.}
    \label{fig:heat_css}
\end{figure}

\begin{figure}[!htb]
    \centering
    \includegraphics[width=0.47\textwidth]{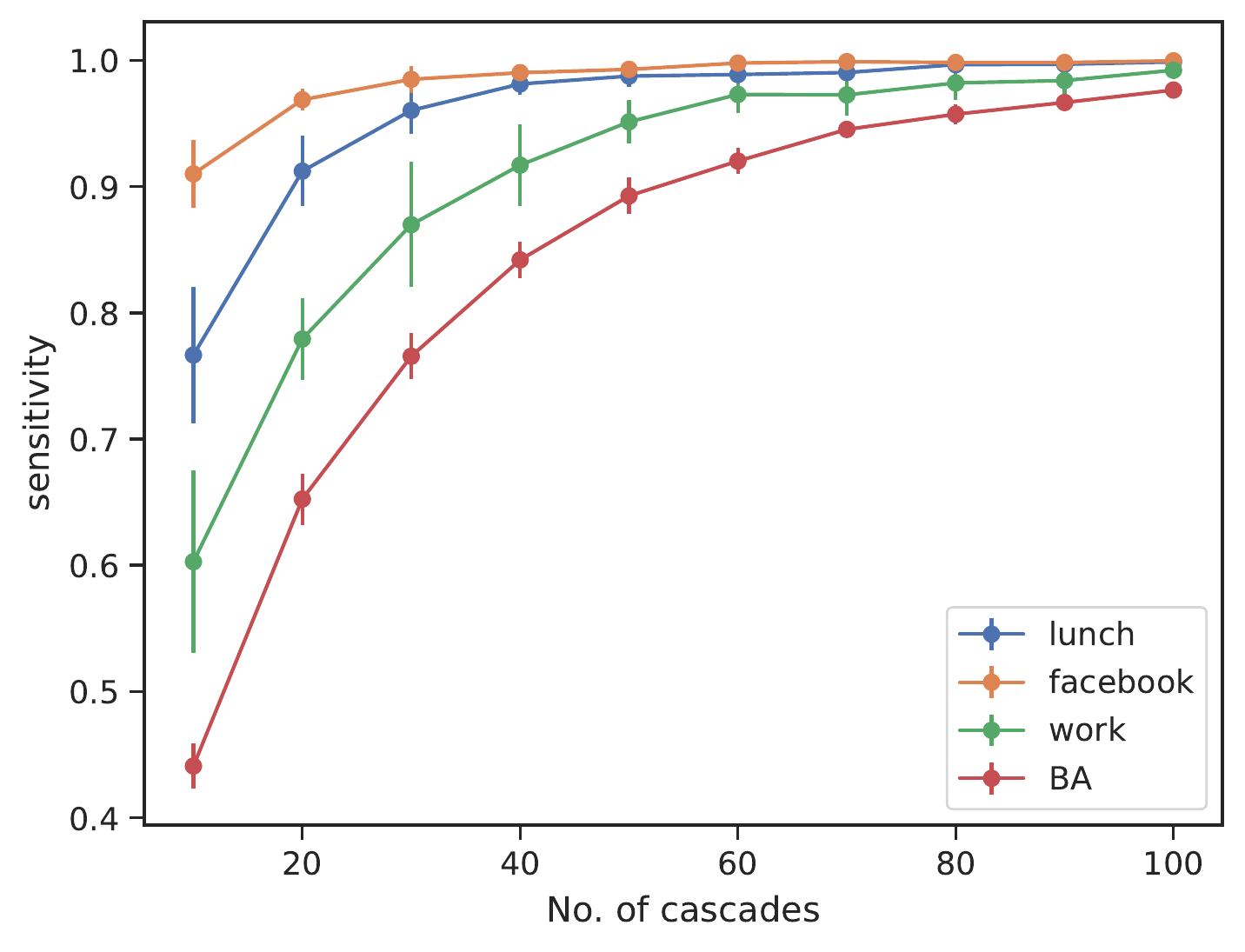}
    \caption{The sensitivity as a function of number of cascades for a) two layer Barabási-Albert network with $m=4$ for both layers, $\beta_{hidden}=0.7$, $\beta_{observed}=0.5$, $N=1000$ and $t_{max}=10$ (red line); b) Aarhus data with different layers as the observed network (\textit{lunch} -- blue line, \textit{facebook} -- yellow line and \textit{work} -- green line).
    The results are averaged over 20 independent runs and the error bars represent one standard deviation.}
    \label{fig:scal}
\end{figure}

As already discussed when the layers are not correlated it is easy to identify that there is a hidden spreading channel.
Nevertheless, finding the actual unobserved links may still be challenging.
Both Fig. \ref{fig:heat_sens} and \ref{fig:heat_css} show that density of the observed network is an important factor.
From the sensitivity perspective it is better to have a denser observed network.
Unfortunately the 0.95-CSS also grows with the density of known connections, making it more demanding to find all the connections.
Additionally, although the effect is weaker, it is beneficial for both measures if the hidden layer is sparser.
This aligns with intuition since more hidden connections can make the observed dynamics much more complex and unsurprisingly having more data about the cascades also makes the task easier.
The actual dependence between the number of cascades and the sensitivity is shown in Fig. \ref{fig:scal}.
For a relatively big BA network we need around 30-40 cascades to reach a fairly satisfactory sensitivity level of around $0.8$.
Depending on the specifics of the problem such amounts of data may be considered a lot (e.g., in epidemic spreading) or easily available (e.g. information spreading on social media).
In the next subsection we will see that scaling for real-life networks.

\subsection{Real World Networks}

On top of synthetic networks we also use real-world data to build a multilayer network and empirically test our methods.
For that purpose we choose the data collected among employees of the Department of Computer Science at Aarhus University \cite{magnani2013combinatorial}.
It is a multilayer network consisting of Facebook friendships, co-authorships, work, leisure (repeated leisure activities) and a lunch layer (regularly eating lunch together).
Its full structure is presented on Fig. \ref{fig:aarhus_struct} with each layer being shown as a separate network. 
The whole network has $61$ nodes and $620$ edges in total.

\begin{figure}[!htb]
    \centering
    \includegraphics[width=0.47\textwidth]{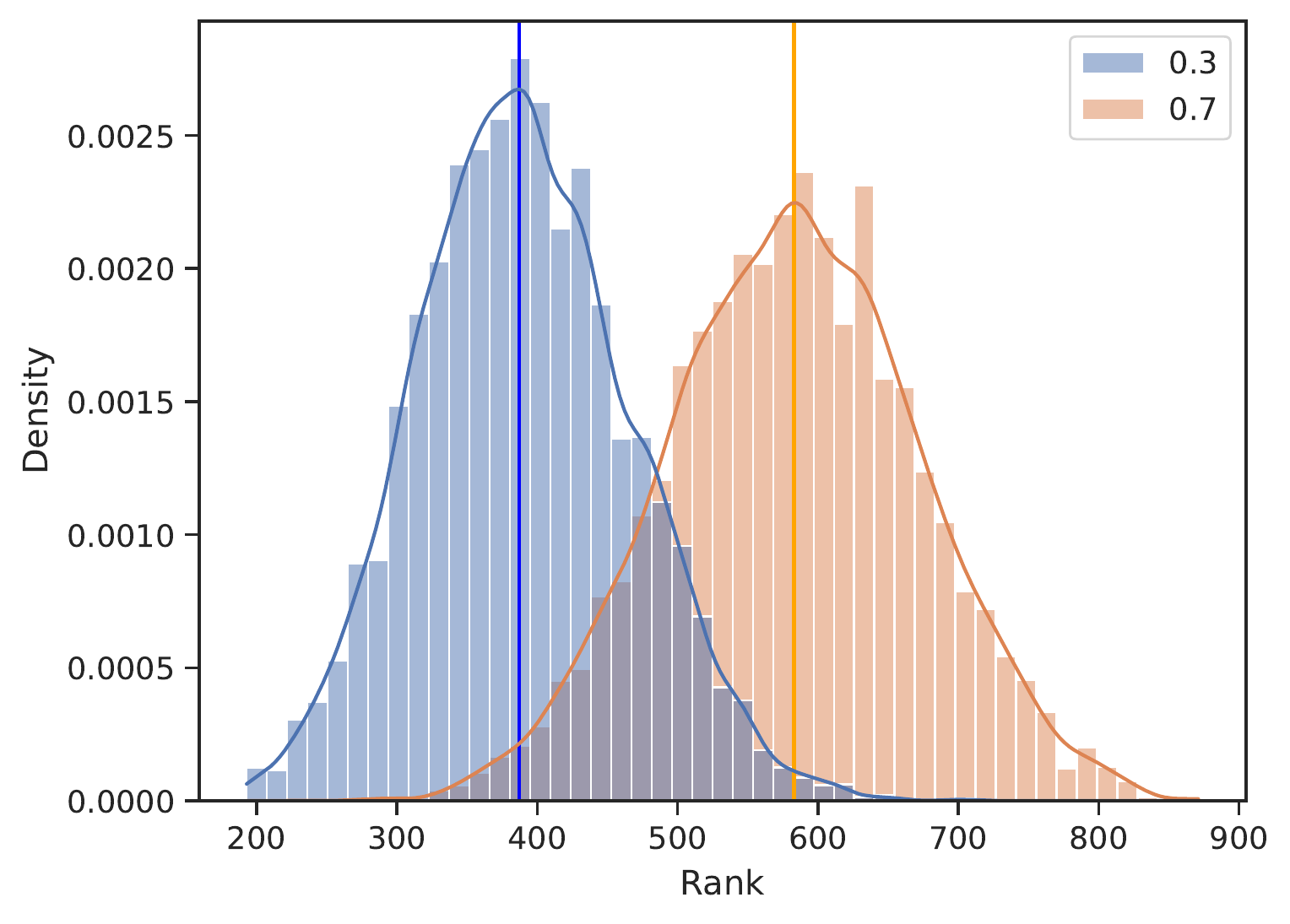}
    \caption{Distribution of ranks of hidden edges, with medians as vertical lines, for the Aarhus data, with the \textit{work} layer as the observed network. Results obtained for 10 cascades with $t_{max}=10$, $\beta_{observed}=0.5$ and two values of $\beta_{hidden}$ -- 0.3 and 0.7. Results from $10^4$ simulations per $\beta_{hidden}$. Solid lines show a Gaussian kernel density estimate.}
    \label{fig:aarhus}
\end{figure}

Main results for the Aarhus data are shown in Tables \ref{tab:facebook_sensitivity}, \ref{tab:work_sensitivity} and \ref{tab:lunch_sensitivity}, which use respectively \textit{facebook}, \textit{work} and \textit{lunch} layers as the visible parts of the graph. 
We omitted the other two possible cases because of their low density of connections.
We treat remaining hidden layers as one, aggregated layer as it does not matter how many layers exactly there are in our detection method.
Sensitivity and CSS values are consistent with synthetic results in the sense that they both grow with the density of the visible layer.
It is also apparent that our approach far exceeds the performance of the null model.
Independently of which layer will be chosen as the visible one, random guessing would require us to check all possible ${61 \choose 2} = 1830$ links, see Eq.~(\ref{eq:null}) and Appendix \ref{sec:app_null}, with $k\in\{159, 160, 229\}$ for \textit{work}, \textit{lunch}, \textit{facebook} as the visible layer respectively.
Our method, on the other hand, needs significantly less than that.
Here, $k$ is the number of unique edges in the whole graph ($353$) minus the number of edges in the visible layer. 
Note that in order to account for the fact that our method does not always find \textit{all} the links, i.e., sensitivity < 1, one can adjust Eq.~\eqref{eq:null} such that one multiplies $k$ by the expected sensitivity.
That, however, barely changes the result, i.e., null model requires maybe one or two edges less than all possible at best. 
Moreover, our method gets more successful the more cascades we can observe.
A more detailed dependence between sensitivity and the number of cascades is shown in Fig. \ref{fig:scal}, where different colors represent different observed layers. 
In Fig. \ref{fig:aarhus} we show the distributions of ranks for the \textit{work} layer as the visible network and when comparing them with the synthetic experiments the two distributions for $\beta_{hidden}=0.3$ and $\beta_{hidden}=0.7$ are much more symmetric and separated.
The distributions of the other two analysed visible layers are qualitatively similar further supporting the merit of our approach (see Appendix \ref{sec:app_aarhus}).

\begin{table}[!htb]
\centering
\begin{tabular}{p{0.08\textwidth}p{0.08\textwidth}p{0.092\textwidth}p{0.092\textwidth}p{0.092\textwidth}}
$\beta_{hidden}$ & $\beta_{observed}$ & \textit{sensitivity} & \textit{0.5-CSS} & \textit{0.95-CSS} \\
\hline
0.3 & 0.5 & 0.92 & 1398 & 1437 \\
0.7 & 0.5 & 0.91 & 1424 & 1475
\end{tabular}
\caption{Sensitivity and $\alpha$-CSS for the Aarhus data, with the \textit{facebook} layer as the observed network. Results obtained for 10 cascades with $t_{max}=10$. Results from $10^4$ simulations per $\beta_{hidden}$.}
\label{tab:facebook_sensitivity}
\end{table}

\begin{table}[!htb]
\centering
\begin{tabular}{p{0.08\textwidth}p{0.08\textwidth}p{0.092\textwidth}p{0.092\textwidth}p{0.092\textwidth}}
$\beta_{hidden}$ & $\beta_{observed}$ & \textit{sensitivity} & \textit{0.5-CSS} & \textit{0.95-CSS} \\
\hline
0.3 & 0.5 & 0.42 & 387 & 515 \\
0.7 & 0.5 & 0.58 & 583 & 729
\end{tabular}
\caption{Sensitivity and $\alpha$-CSS for the Aarhus data, with the \textit{work} layer as the observed network. Results obtained for 10 cascades with $t_{max}=10$. Results from $10^4$ simulations per $\beta_{hidden}$.}
\label{tab:work_sensitivity}
\end{table}

\begin{table}[!htb]
\centering
\begin{tabular}{p{0.08\textwidth}p{0.08\textwidth}p{0.092\textwidth}p{0.092\textwidth}p{0.092\textwidth}}
$\beta_{hidden}$ & $\beta_{observed}$ & \textit{sensitivity} & \textit{0.5-CSS} & \textit{0.95-CSS} \\
\hline
0.3 & 0.5 & 0.71 & 732 & 838 \\
0.7 & 0.5 & 0.78 & 841 & 966
\end{tabular}
\caption{Sensitivity and $\alpha$-CSS for the Aarhus data, with the \textit{lunch} layer as the observed network. Results obtained for 10 cascades with $t_{max}=10$. Results from $10^4$ simulations per $\beta_{hidden}$.}
\label{tab:lunch_sensitivity}
\end{table}

\section{Discussion}

Spreading processes on networks are a valuable tool when describing real-life global diffusion processes, like epidemics, information spreading, cascading failures etc.
These processes may have several spreading channels and rarely do we know, or are even aware, of all of them.
It is therefore crucial to identify whether observed spreading was in fact generated only by the observed network.
Furthermore, should one confirm the existence of an unobserved spreading path, finding these hidden connections can be of the utmost importance.
    
In this paper we focused on identifying both the existence and the structure of a hidden spreading layer by observing a diffusion process unraveling on a graph. We provide methods for i) determining whether a hidden layer exists and ii) estimating what links are present in that layer.
Our approach is based on an exact formula for the likelihoods of an observed cascade given knowledge of the system's topology. Using said likelihood and the fact its distribution can be assumed to be unimodal we established a practical and effective way of discerning the existence of a hidden layer.
Furthermore using a series of heuristics we obtain an algorithm for estimating the joint likelihood of given (hidden) edge taking part in the observed cascade therefore providing a tool for assessing which nodes are most likely to exchange information via channel we do not know of that is vastly superior to random guessing.
In short, a typical situation where our approach could be used is when we observe a single network and a spreading process described by a known model and we suspect that there are hidden layers through which the spread may progress.
Note that the observed network can already have multiple layers and the hidden part can also consist of more than one layer.
Finally, one could also use our method in a setting where only one layer exist, but at the same time some links are not observed.

Data from synthetic and empirical networks alike confirm that uncovering the hidden spreading channel is a relatively simple task with our approach - especially when the layers are uncorrelated.
It is, however, more difficult to identify specific hidden connections.
Despite the general similarities there are some quantitative differences between the results obtained with synthetic and real data.
One of the most significant differences is how $\beta_{hidden}$ relates to the distribution of ranks of hidden edges, influencing the difficulty of hidden connections reconstruction.
This effect is much stronger in real world networks than in synthetic ones.
It can, however, be explained by the difference in density between hidden and observed networks.
In the corresponding plots for synthetic data (see Fig. \ref{fig:ranks}) both layers have the same density.
Here the hidden layer is denser (it is a sum of four hidden layers) and so changing the hidden spreading probability affects majority of connections.

An important factor in being able to successfully recover the hidden connections turns out to be the density of both the hidden and the observed transmission layers.
Specifically, we observe that the denser the hidden layer the harder it is to find the exact connections.
An interesting interplay takes place when it comes to the density of the observed layer.
On one hand the sensitivity decreases with the density of observed layer, on the other hand, the $\alpha$-CSS is also decreasing with the density.
This observation, confirmed by both synthetic and real data, means that as the number of connections on a visible layer increases, we are able to identify less hidden edges on average but we need to take into account a smaller set of potential edges in order to find all of the hidden connections.

It should be pointed out that we only focus on the hidden connections which are not overlapping with the observed ones.
This means that for correlated layers there might be only few unknown connections whereas the overlapping edges are also influencing the dynamics.
Focusing on the more general picture and including the overlapping connections is an interesting subject for future research.
Another research direction would be to focus on further improving the hidden connections identification algorithms.
These improvements should include both the effectiveness and scalability of proposed methods.
The latter is specifically important since real world networks are often quite substantial in size.
From the perspective of empirical data it would also be useful to have a way of handling a scenario where different layers have different values of $\beta$ which may or may not be known.
Finally, a more radical generalisations like including temporal networks could also prove to be an interesting research problem.
While we do hope to address some of the above topics in the near future we feel that methods presented here already provide effective and practical tools for real world applications.

\acknowledgements{Ł.G.G and J.C. were supported by National Science Centre, Poland Grant No. 2015/19/B/ST6/02612.
M.W. acknowledges support from the Laboratory Directed Research and Development program of Los Alamos National Laboratory under projects numbers 20200121ER and 20210529CR.}

\bibliographystyle{unsrt}  
\bibliography{references}

\appendix

\section{Independent-Cascade model}
\label{sec:app_ic}

\begin{figure*}[htb]
    \centering
    \includegraphics[width=0.47\textwidth]{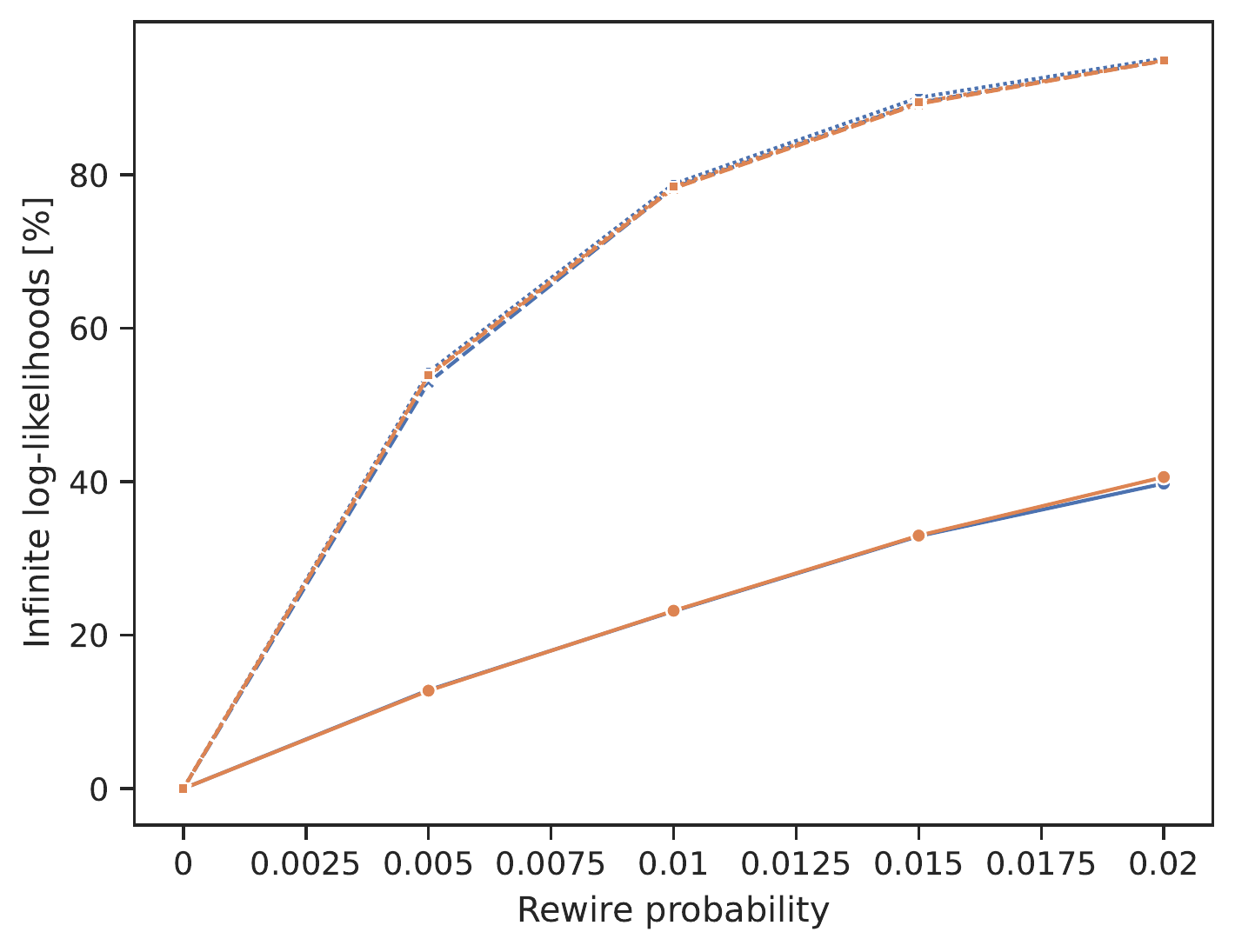}
    \hfill
    \includegraphics[width=0.47\textwidth]{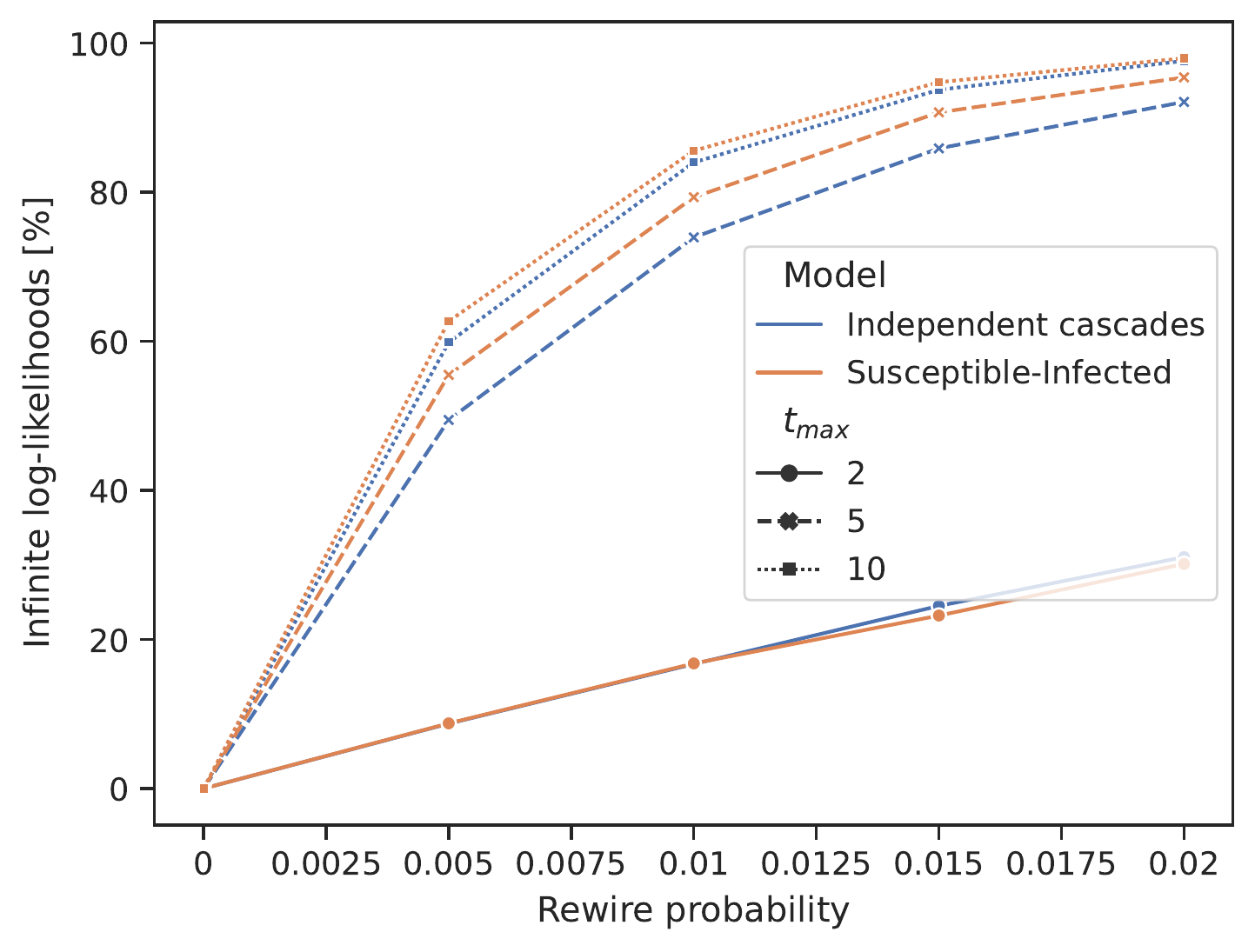}
    \caption{The percentage of log-likelihoods resulting with $-\infty$ as a function of the probability of rewiring $p$. Simulations were conducted using two different spreading models - Susceptible Infected (presented in main text) and Independent Cascades for comparison.
    We investigate two different cases of networks: a square lattice (right) and the the Barabási-Albert network (left) for different lengths of cascades.
    The simulations were made for networks of size $N=100$ with periodic boundary conditions in the lattice case.}
    \label{fig:app_infty}
\end{figure*}

\begin{figure*}[htb]
    \centering
    \includegraphics[width=0.47\textwidth]{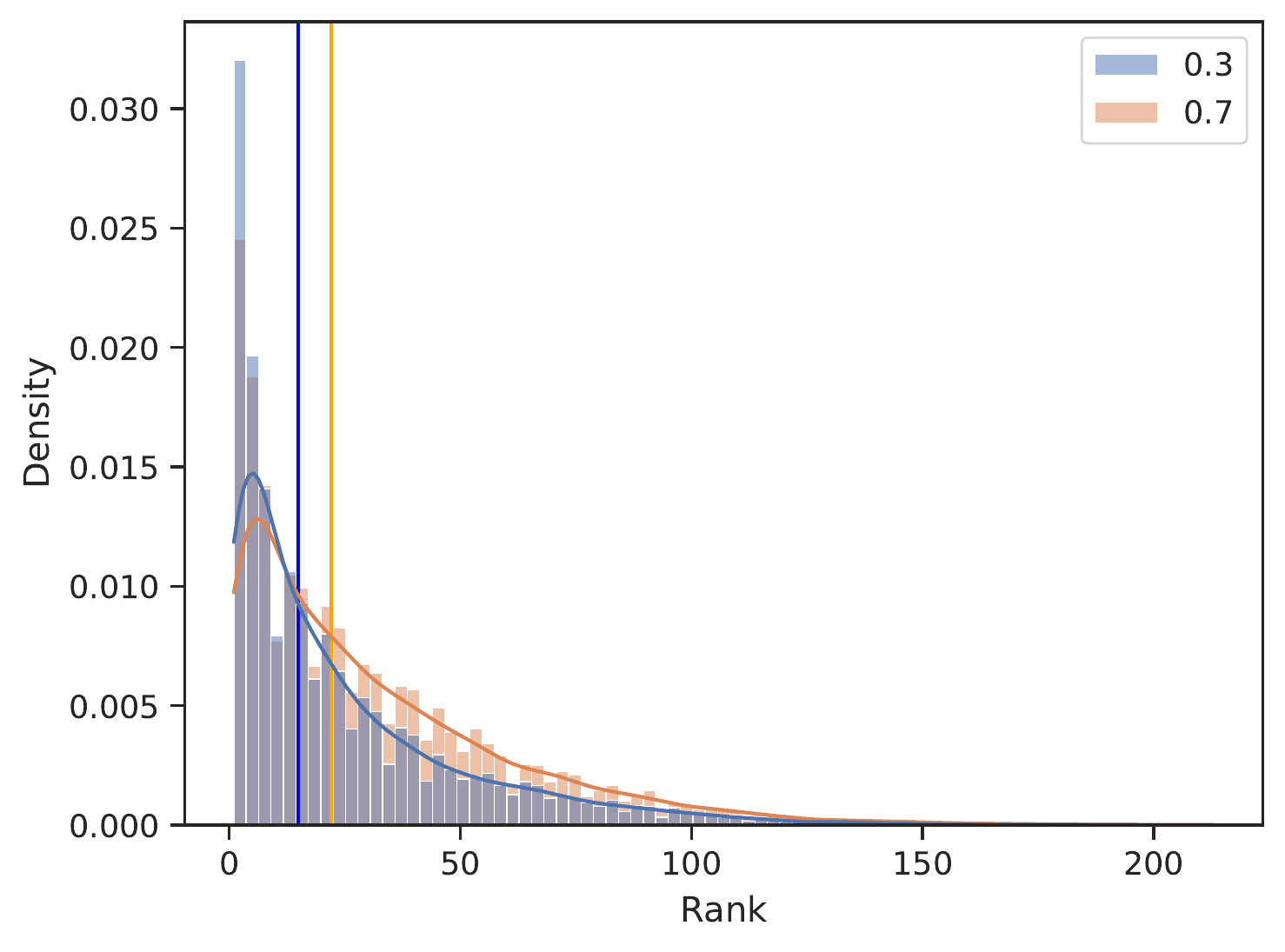}
    \hfill
    \includegraphics[width=0.47\textwidth]{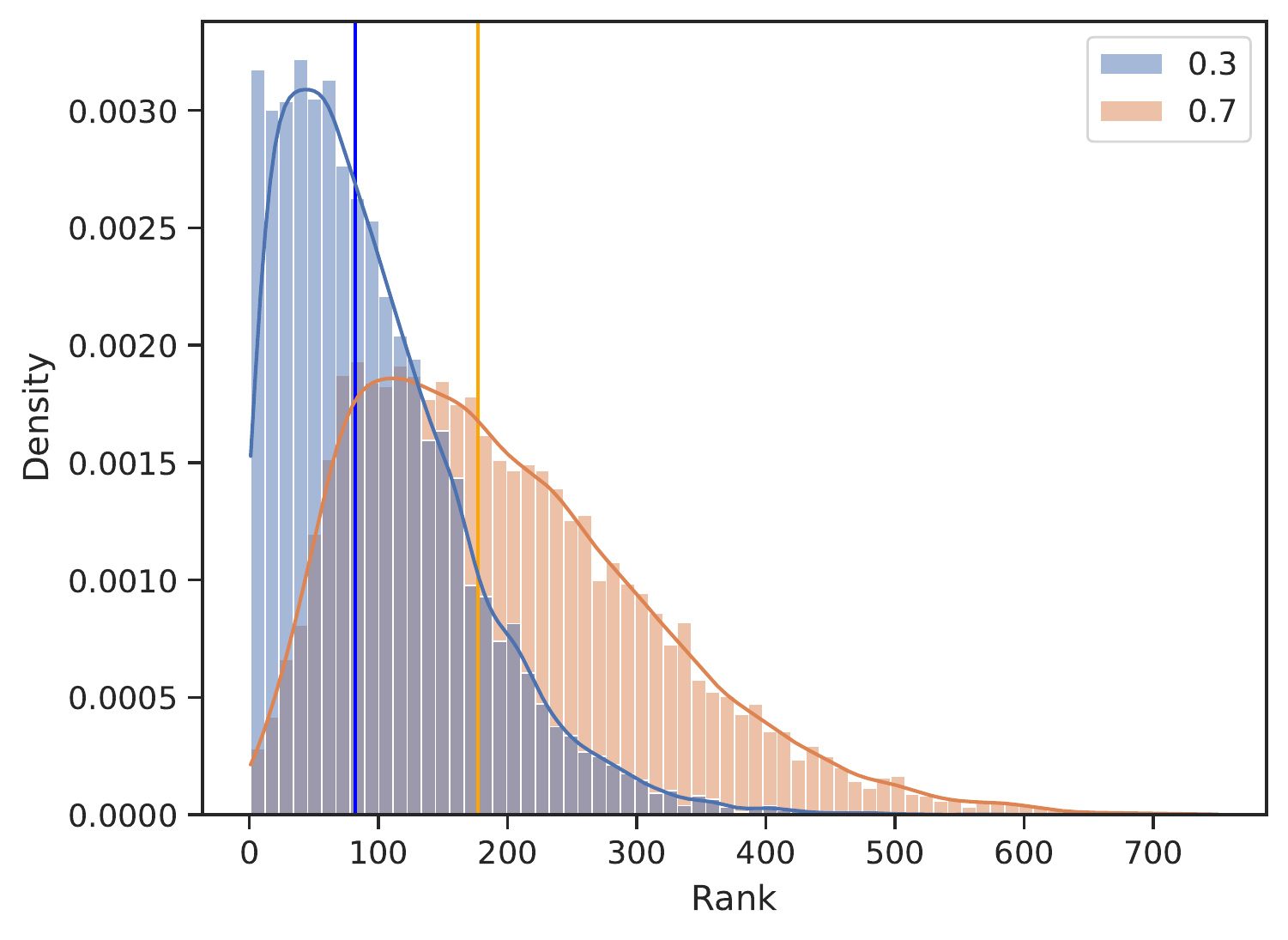}
    \caption{Distribution of ranks of hidden edges with medians as vertical lines. Simulations were conducted using the Independent Cascade model for comparison against the results from the main text (which show results for the Susceptible Infected model).
    Right: lattice with rewiring ($N=100$, $t_{max}=10$, $p=0.01$). Left: BA network with rewiring ($N=100$, $t_{max}=10$, $m=3$, $p=0.01$).
    Results obtained with $10^4$ realisations per scenario where each scenario had $10$ independent cascades from different sources.
    These results are for those realisations where all hidden edges were detected. Solid lines show a Gaussian kernel density estimate.}
    \label{fig:app_ranks}
\end{figure*}

As mentioned in the main text, our methodology can easily be applied to different spreading processes.
One example of such process is the IC model.
The factorised form of the likelihood, in case of the IC model, is the same as for the Susceptible-Infected model
\begin{equation}
    P(\Sigma | G, \{ \beta_j \}) = \prod_{i \in V} \prod_{c \in C} P_i(\tau^c_i | \Sigma^c, G, \{ \beta_j \}).
    \label{eq:ic_likelihood}
\end{equation}
The form of the local probability, however, is different
\begin{equation}
    \begin{split}
        &P_i(\tau_i^c | \Sigma^c, G, \{ \beta_j \}) = \Bigg(\prod_j \prod_{k \in \partial_j i} (1 - \beta_j \mathbf{1}_{\tau_k^c \leq \tau_i^c - 2}) \Bigg) \\
        &\times \Bigg( 1 - \prod_j \prod_{k \in \partial_j i} (1 - \beta_j \mathbf{1}_{\tau_k^c = \tau_i^c - 1}) \mathbf{1}_{\tau_i^c < t_{max}} \Bigg),
        \label{eq:ic_local_likelihood}
    \end{split}
\end{equation}
which is only valid for $\tau_i^c > 0$, otherwise it is equal to $1$.
The plots equivalent to Fig. \ref{fig:infinite} and Fig. \ref{fig:ranks} in the case of the IC model are shown respectively in Fig. \ref{fig:app_infty} and Fig. \ref{fig:app_ranks}.
Note that these results do not deviate significantly from the SI simulations.
This is interesting for two reasons.
First, one could easily produce artificial examples where either IC or SI would be very easier to recover.
On one hand it should be easier for IC model to spot any forbidden dynamics, because of the restrictive condition regarding only one time step where spreading is possible for a single node.
SI, on the other hand, ensures more statistics, because of the opposite situation (nodes can spread at any time once they are infected).
Second, since these are limiting versions of the SIR model, all other variations (where recovery probability is between $0$ and $+\infty$) will also produce similar results.
This makes the shown results much more universal, than just for the two analysed models.

\begin{figure}[htb]
    \centering
    \includegraphics[width=0.47\textwidth]{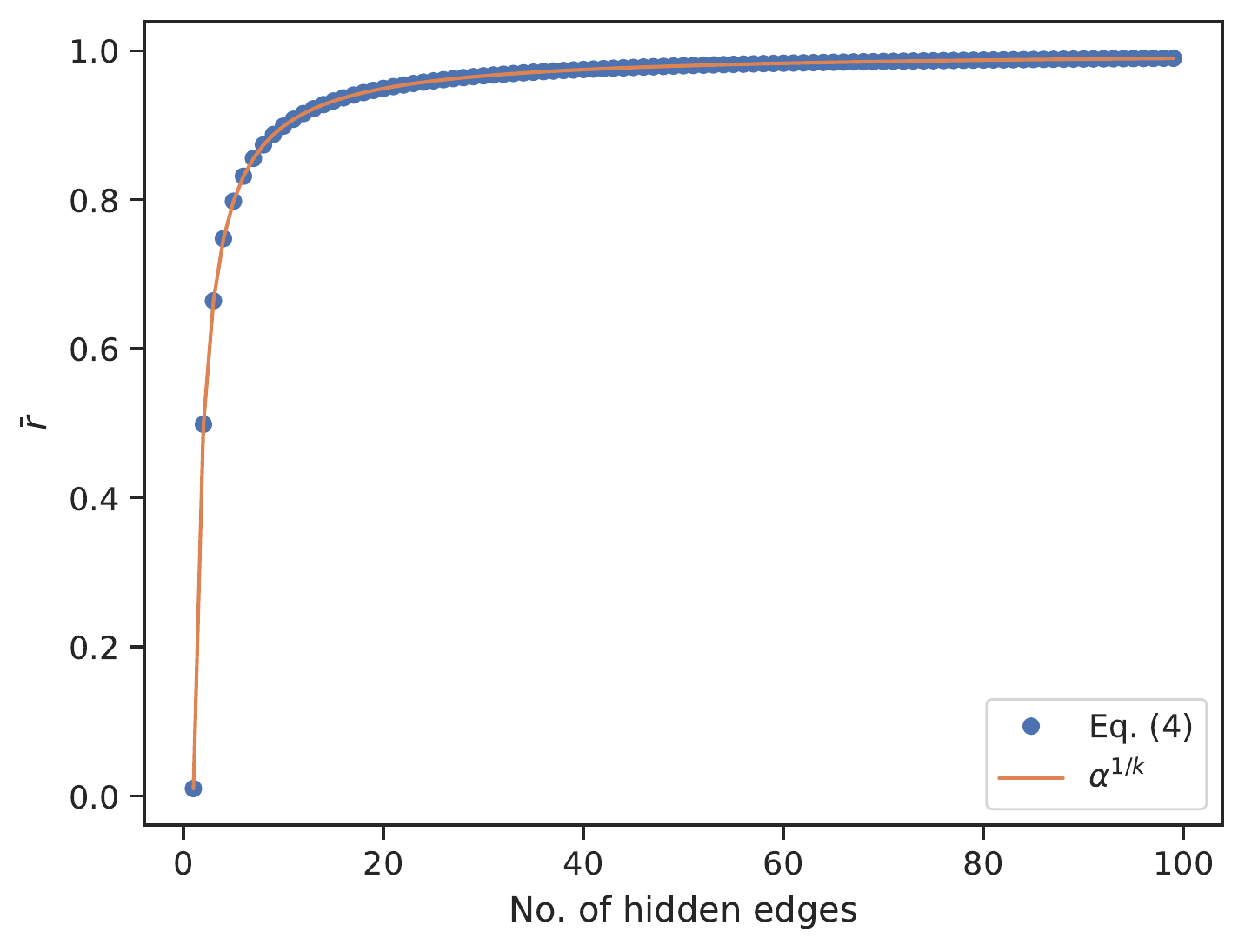}
    \caption{Comparison of numerical solution of Eq.~(\ref{eq:null}) and the approximation (\ref{eq:asympt}) for $N=100$. The curve represents the fraction of edges required to check by the null model in order to have 95\% certainty of testing all hidden edges.}
    \label{fig:null_model_asympt}
\end{figure}

\section{Null model}
\label{sec:app_null}
In the main text we use Eq.~(\ref{eq:null}) as our null model, however, that form must (for the most part) be solved numerically for $r$ with a set $\alpha$. Here we present an approximation that gives a closed form for $r$ and provides an excellent match to numerical computations (and is of course much easier and faster to compute).

Let us recall the aforementioned formula:
\begin{equation} 
\alpha=\frac{{r \choose k}}{{{N \choose 2} \choose k}} = \frac{r!}{\left(r-k\right)!}\frac{\left({N \choose 2}-k\right)!}{{N \choose 2}!}.
\end{equation}
Denote $N \choose 2$ as $\xi$, then
\begin{equation}
\begin{aligned}
    \alpha &= \frac{r!}{(r-k!)}\frac{(\xi-k)!}{\xi!} \\
    &= \frac{(r-1)(r-2)\dots(r-k)!}{(r-k!)}\frac{(\xi-k)!}{(\xi-1)(\xi-2)\dots(\xi-k)!} \\
    &= \frac{(r-1)(r-2)\dots(r-k+1)}{(\xi-1)(\xi-2)\dots(\xi-k+1)} \sim \left(\frac{r}{\xi}\right)^k.
\end{aligned}
\end{equation}
We can therefore conclude that
\begin{equation}
    r \sim \frac{1}{2} N (N-1) \sqrt[k]{\alpha}.
    \label{eq:asympt}
\end{equation}
This leads to:
\begin{equation}
    \bar{r} = \frac{2r}{N(N-1)} \sim \sqrt[k]{\alpha}.
    \label{eq:asympt_r_bar}
\end{equation}
The comparison of this result to the numerical solution is shown in Fig.~\ref{fig:null_model_asympt}.

\section{Distributions of ranks for Aarhus data}
\label{sec:app_aarhus}
\begin{figure*}[htb]
    \centering
    \includegraphics[width=0.47\textwidth]{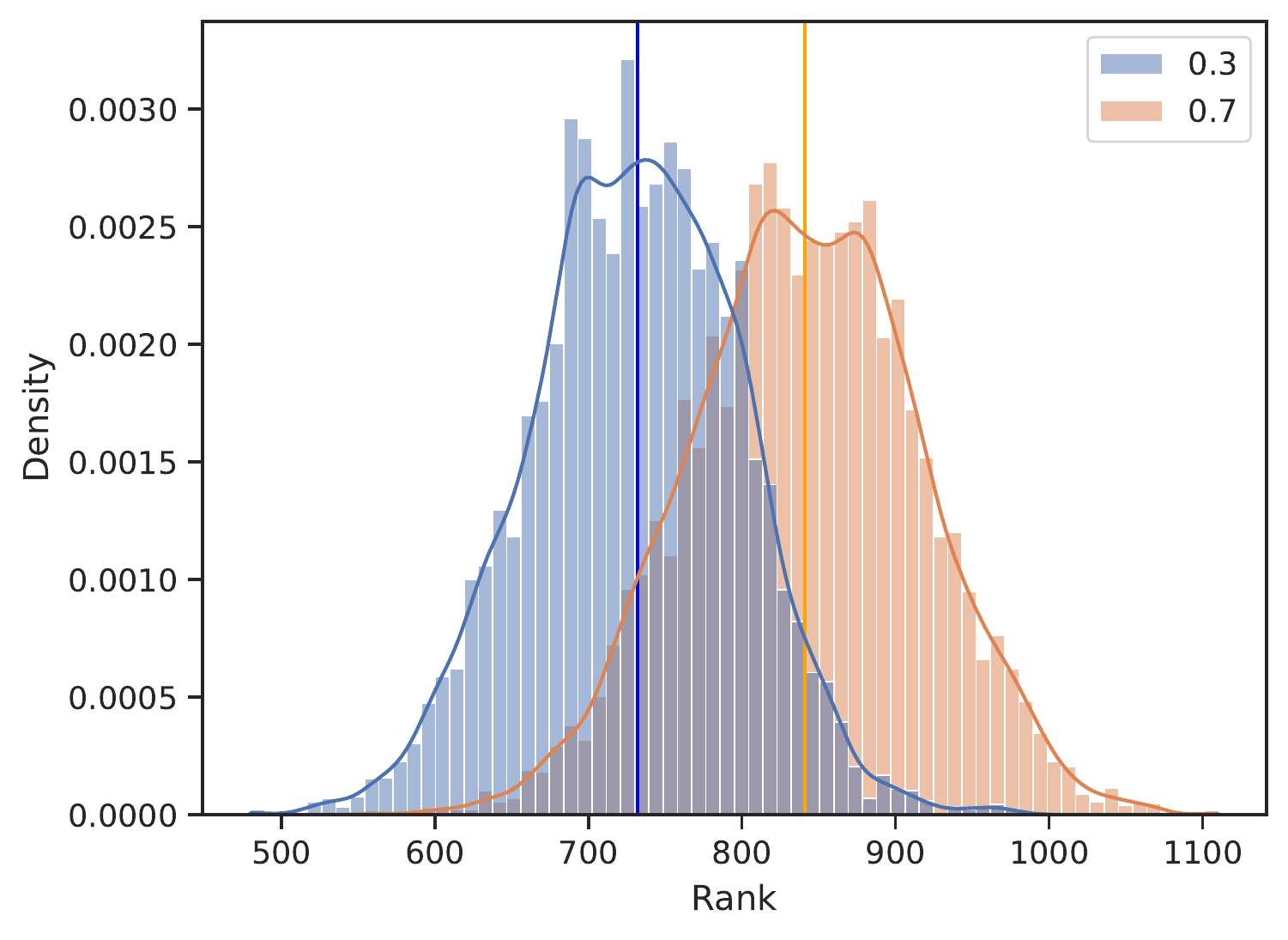}
    \hfill
    \includegraphics[width=0.47\textwidth]{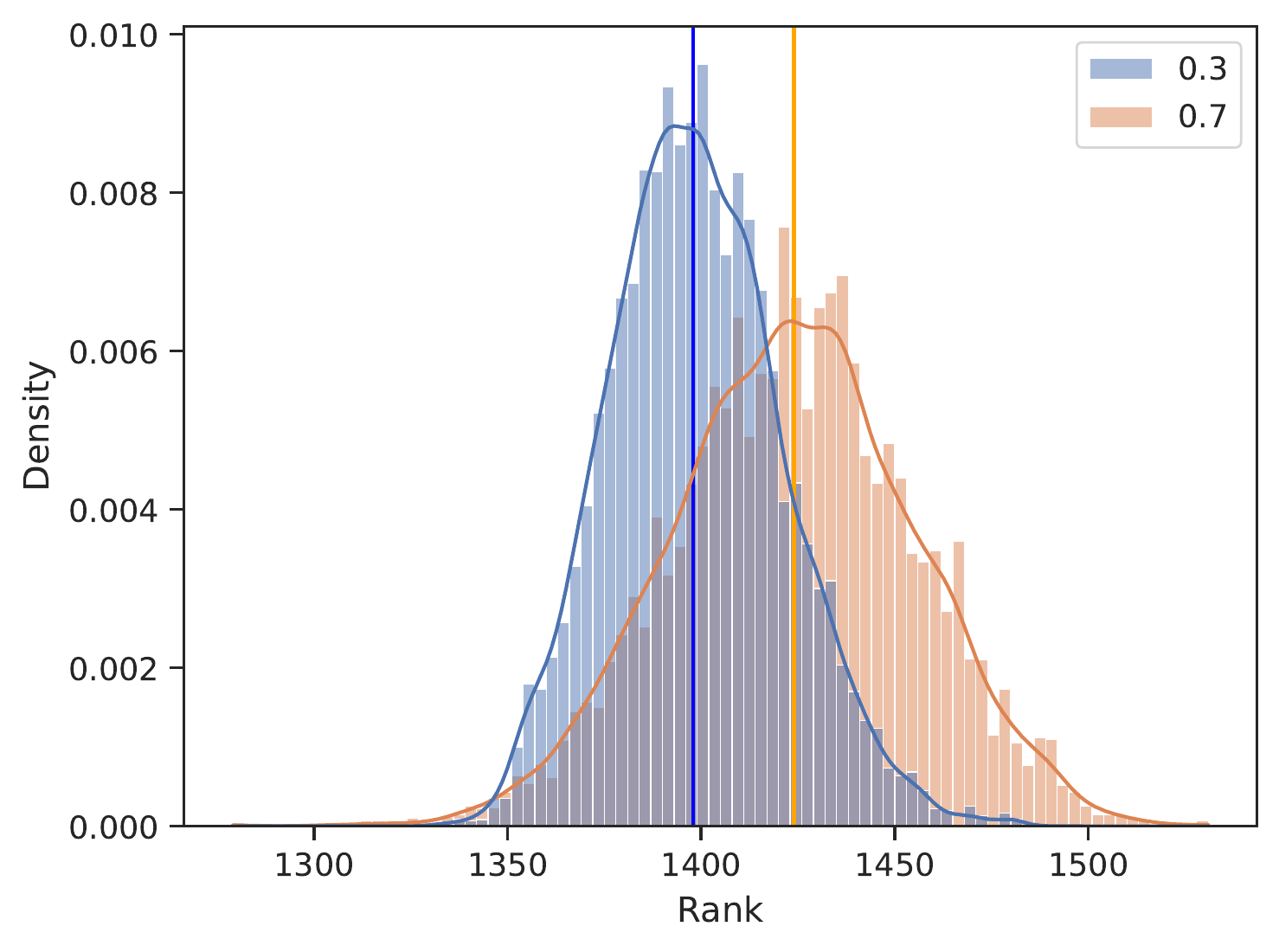}
    \caption{Distribution of ranks of hidden edges, with medians as vertical lines, for the Aarhus data, with the \textit{lunch} (left) and \textit{facebook} (right) layer as the observed network. Results obtained for 10 cascades with $t_{max}=10$, $\beta_{observed}=0.5$ and two values of $\beta_{hidden}$ -- 0.3 and 0.7. Results from $10^4$ simulations per $\beta_{hidden}$. Solid lines show a Gaussian kernel density estimate.}
    \label{fig:aarhus_lunch_facebook}
\end{figure*}

We analyse three scenarios for the Aarhus data (described in the main text).
For each one of them we use different layer as the visible one and project all the other layers to one hidden network.
The distributions of ranks obtained by applying our approach to the \textit{lunch} and \textit{facebook} as observed layer scenarios, are shown in Fig. \ref{fig:aarhus_lunch_facebook}.
The third scenario is shown in Fig. \ref{fig:aarhus}, in the main text.

\section{Layer detection dependence on $\beta$ and system size}
\label{sec:app_p_of}

We test how our layer detection scheme holds for different infection rates and system sizes. By conducting analogous experiments to those presented in Fig.~\ref{fig:p_max} we can show the mean and standard deviation of $p(\tilde{x})$ from Eq.~(\ref{eq:p_max}) as a function of these variables. We set the visible infection rate $\beta_{visible} = 0.5$, $t_{max}=10$, and simulate $10^4$ times per data point. 

\begin{figure*}[htb]
    \centering
    \includegraphics[width=0.60\textwidth]{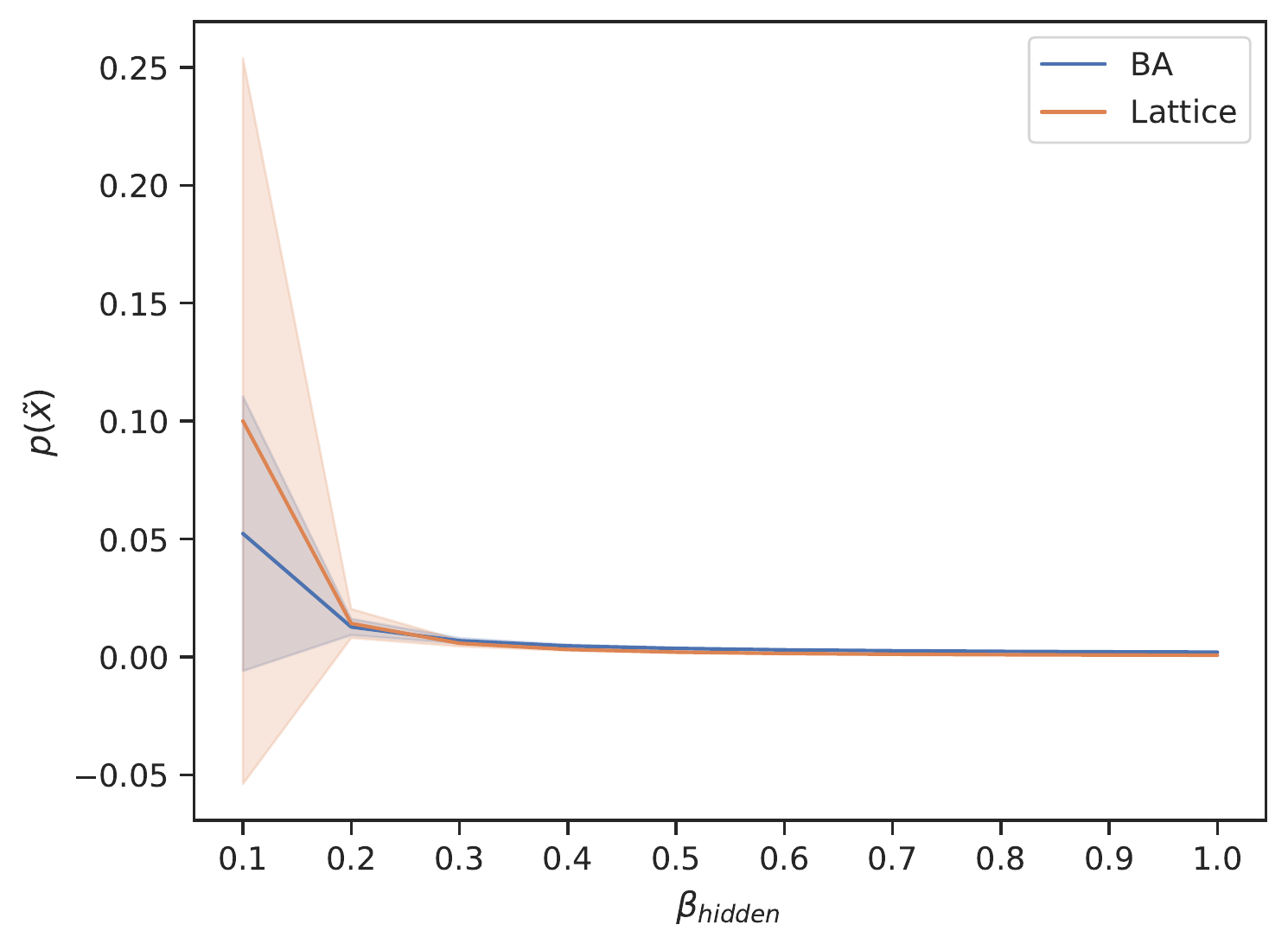}
    \caption{Mean probability that the system has only one layer as a function of the infection rate on the hidden layer - $\beta_{hidden}$ - see Eq.~\ref{eq:p_max} and Fig.~\ref{fig:p_max}. The colour bands represent one standard deviation. Results for two graphs are presented - BA ($m=3$) and a square lattice with system size $N=100$, $\beta_{visible}=0.5$, and $t_{max}=10$. Each point is the results of $10^4$ simulations.}
    \label{fig:app_p_of_beta}
\end{figure*}

In Fig.~\ref{fig:app_p_of_beta} we show results for a BA graph ($m=3$) and a square lattice with the infection rate of the hidden layer $\beta_{hidden} \in [0.1, 1.0]$. When $\beta_{hidden}$ is small we see a wide spread of the $p(\tilde{x})$ values which is not surprising as the probability of a hidden edge being the preferred connection is also small. As the hidden infection rate rises both the expected value and the variance diminish quickly showing that if the hidden layer is significant enough to cause an effect in the whole system then our method will detect it.

\begin{figure*}[htb]
    \centering
    \includegraphics[width=0.47\textwidth]{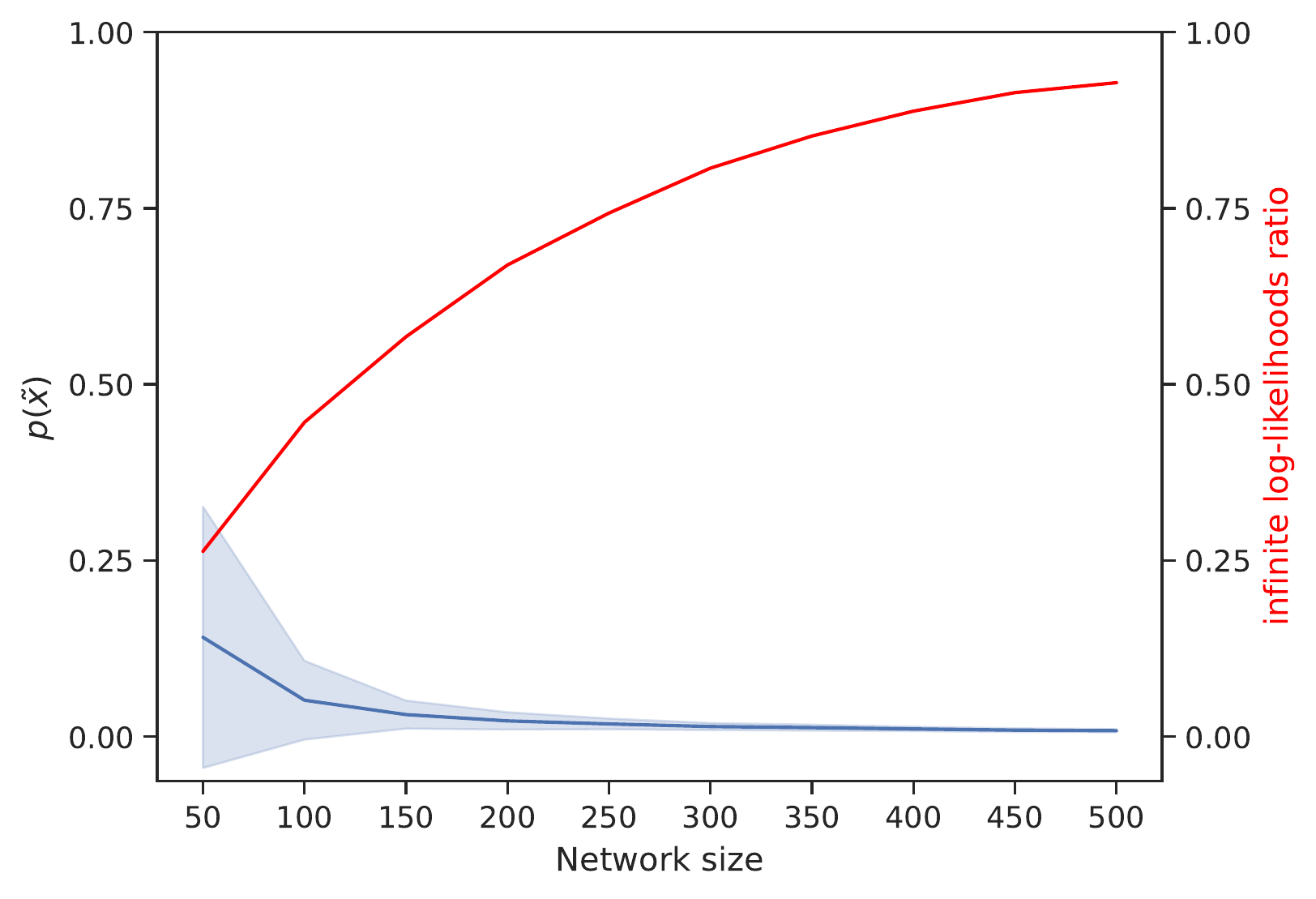}
    \hfill
    \includegraphics[width=0.47\textwidth]{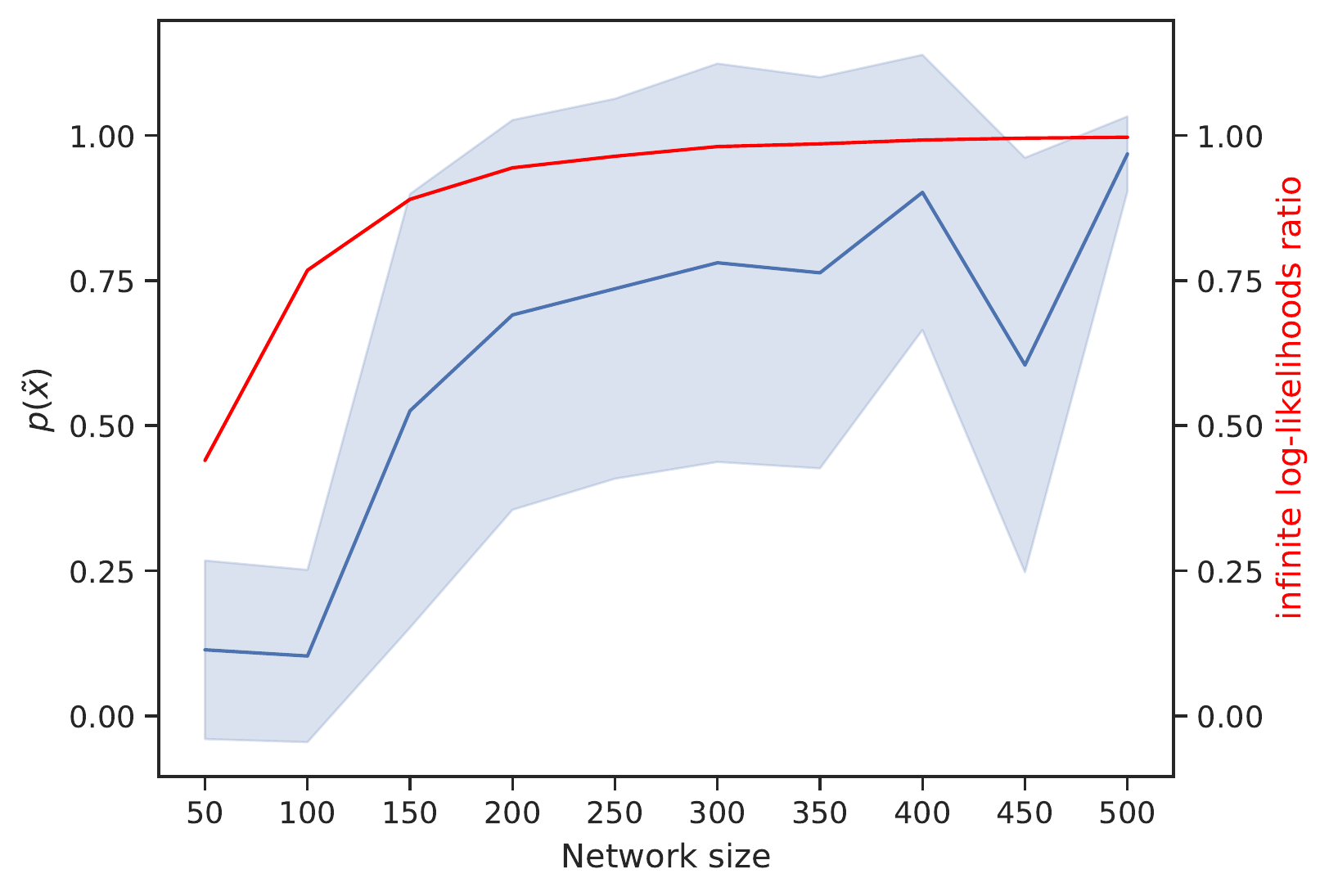}
    \caption{Mean probability that the system has only one layer as a function of the system size (blue) - see Eq.~\ref{eq:p_max} and Fig.~\ref{fig:p_max}. The hidden layer infection rate - $\beta_{hidden} = 0.1$. The colour bands represent one standard deviation. Results for two graphs are presented - BA ($m=3$, left) and a square lattice (right) $\beta_{visible}=0.5$, and $t_{max}=10$. Each point is the results of $10^4$ simulations. The red curves (Y-axis values on the right side of the plots) represent the ratio of infinite ($-\infty$) log-likelihoods detected in the data (see Fig.~\ref{fig:infinite}). The wide spread and increasing behaviour for the lattice is the result of a very high infinite log-likelihoods ratio - by the time we reach $N=500$ only an order of 30 realisations out of $10^4$ are finite.}
    \label{fig:app_p_of_n}
\end{figure*}

In Fig.~\ref{fig:app_p_of_n} we show how the size of the network, $N \in [50, 500]$, affects the $p(\tilde{x})$. On the left we present results for BA ($m=3$) and on the right for a square lattice. In both cases we choose $\beta_{hidden} = 0.1$ as this is the ``worst case'' scenario from the previous plot. The blue line shows the mean and one standard deviation of $p(\tilde{x})$ while the red shows the ratio of infinite log-likelihoods. For the BA graph the statistic is still sufficient in the tested range, however, for the lattice it is not. On the left we can see that as the size increases the mean and variance diminish which is due to the fact that the larger the graph the more potentially ``forbidden'' links are possible, and thus the detection of the hidden layer becomes easier. On the right the blue curve shows some wild values and that is because of how quickly the red curve rises up to virtually being equal 1. By the time we reach $N=500$ only an order of 30 realisations out of $10^4$ are finite, rendering the blue curve a bit misrepresentative as the statistic is insufficient and biased towards cascades very similar to those generated by a single layer model. We show it here for the sake of completeness.

\end{document}